\documentclass{WileyMSP-template}
\usepackage[utf8]{inputenc}
\usepackage{graphicx}
\usepackage{textcomp}
\usepackage{txfonts}
\usepackage{url}
\urldef{\SMonline}\url{https://journals.aps.org/prb}
\usepackage{color}
\usepackage{gensymb} % ° degree symbol
\usepackage{tikz}
\usetikzlibrary{positioning} % arrows in math
\usetikzlibrary{calc} % arrows in math
\usetikzlibrary{shapes}
\usetikzlibrary{arrows}

\renewcommand{\vec}[1]{{\bf #1}}

\newcommand{\jena}{Institute of Physical Chemistry and Institute of Condensed Matter Theory and Optics, Friedrich Schiller University, F\"{u}rstengraben 1, 07743 Jena, Germany}

\newcommand{\new}[1]{\textcolor{black}{#1}}

\begin{document}

\pagestyle{fancy}
%\rhead{\includegraphics[width=2.5cm]{vch-logo.png}}
\rhead{}

%\title{Electronic transport at ferroelectric domain walls in $n$-doped BiFeO$_3$}
%\title{Energy barriers for electron hopping in bismuth ferrite from first principles}
\title{\new{Energy profile and hopping barriers for small electron polarons at ferroelectric domain walls} in bismuth ferrite from first principles}
%\title{Activation energy for electronic transport in bismuth ferrite from first principles}

\maketitle

% Author: Please give full first and last names for authors and include * after the name of all corresponding authors

\author{Sabine K\"{o}rbel*}

% Dedication

\dedication{}

% Affiliations: Please provide adacemic titles (Prof. or Dr.) for all authors where applicable, and include an institutional email address for all corresponding authors
\begin{affiliations}
\jena \\
Email Address: skoerbel@uni-muenster.de
\end{affiliations}

% Keywords: Please provide a minimum of three and a maximum of seven keywords, separated by commas


% Abstract should be written in the present tense and impersonal style (i.e., avoid we), and be at most 200 words long
\begin{abstract}
Evidence from first-principles calculations indicates that excess electrons in BiFeO$_3$ form small polarons
with energy levels deep inside the electronic band gap.
Hence, $n$-type electronic transport could occur by hopping of small electron polarons rather than by band-like transport.
Here, by means of first-principles calculations, small electron polaron hopping in BiFeO$_3$ is investigated.
Both bulk BiFeO$_3$ and a typical ferroelectric domain wall, the neutral 71\textdegree~domain wall, are considered. 
The latter is included to account for experimental observations of electrical conductivity at domain walls in otherwise insulating ferroelectrics. 
The object of this study is to shed light on the intrinsic electron conduction in rhombohedral BiFeO$_3$ and the effect of pristine neutral ferroelectric domain walls.
The computed energy barriers for small electron polaron hopping are \new{near 0.2\,eV, similar to other perovskite oxides}, 
both in the bulk and within the neutral 71\textdegree~domain wall.
\new{Trapping energies of small electron polarons at the three prevalent domain walls, 
the 71\textdegree, the 109\textdegree, and the 180\textdegree\ wall, were determined.}
The domain walls are found to act as two-dimensional traps for small electron polarons, with a trap depth of about two times the thermal energy at room temperature.
Based on these findings, the intrinsic $n$-type mobility and the diffusion constant in BiFeO$_3$ at room temperature are estimated, 
and experimental conductivity data for BiFeO$_3$ are discussed.
\end{abstract}
%%%%%%%%%%%%%%%%%%%%%%%%%%%%%%%%%%%%%%%%%%%%%%%%%%%%%
\section{Introduction}
\begin{figure}[htb]
	 \includegraphics[width=0.45\textwidth]{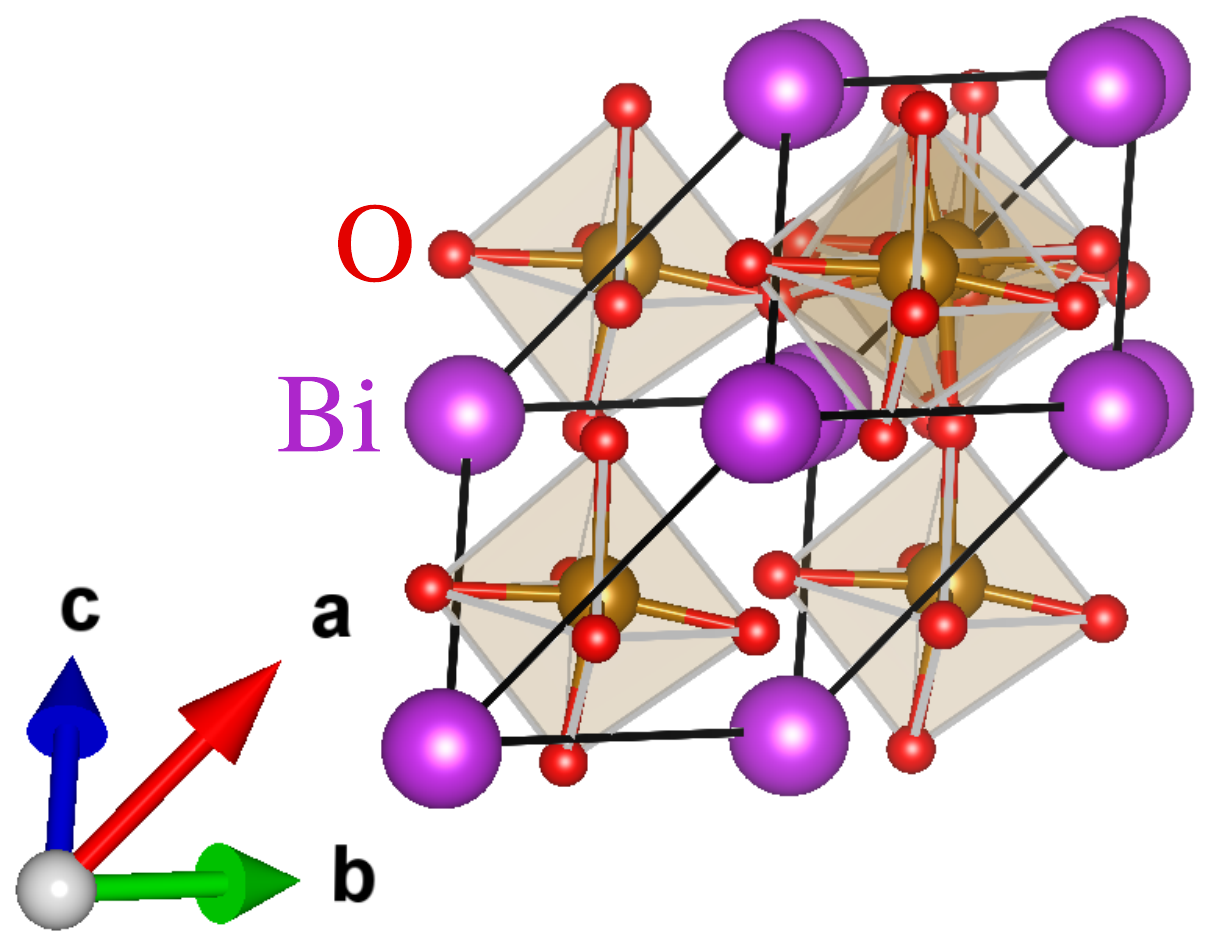}
	 \hspace{0.1\textwidth}
	\includegraphics[width=0.3\textwidth]{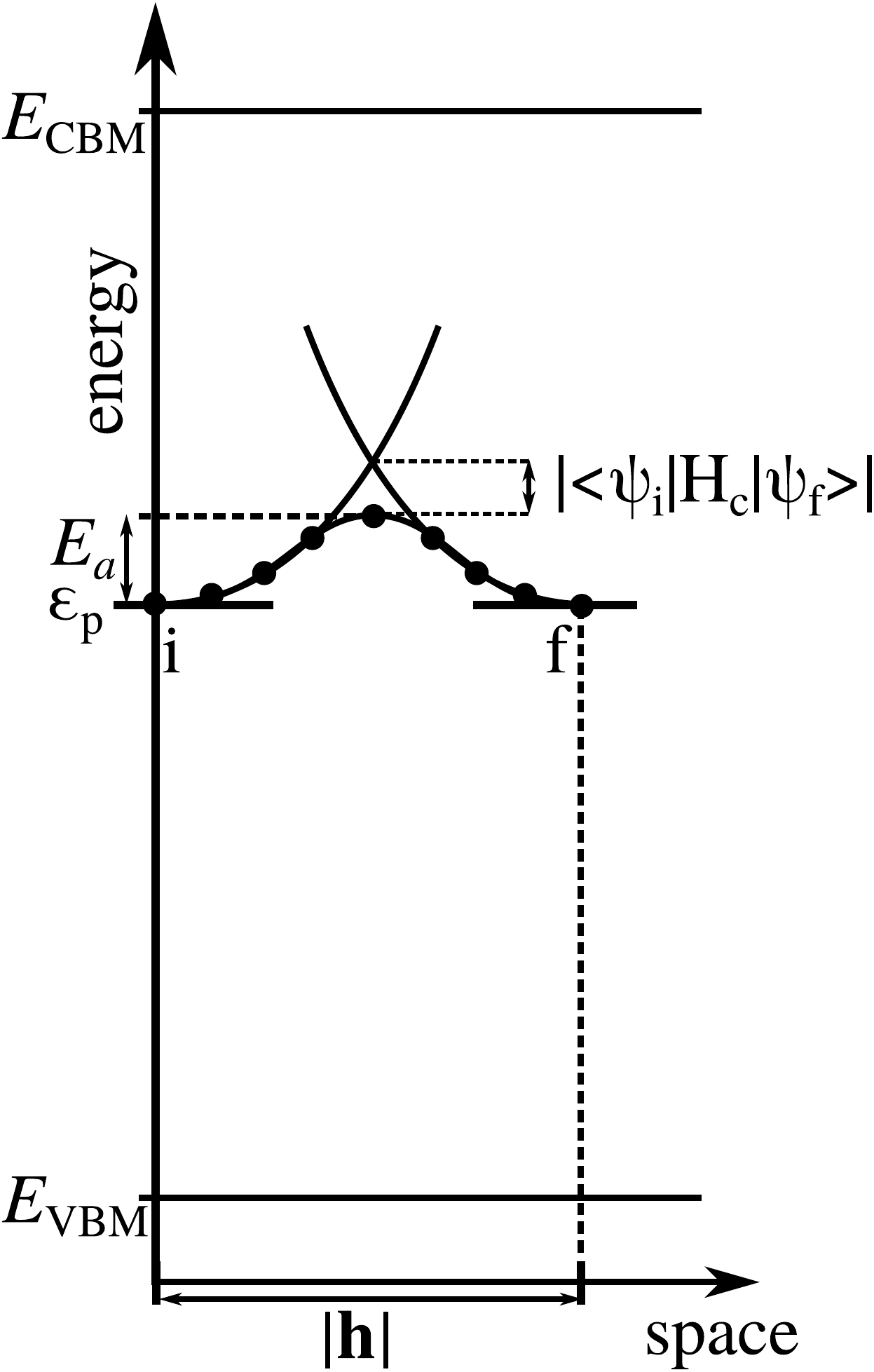}
	 \put(-450,180){(a)}
	 \put(-180,180){(b)}
	 \caption{\label{fig:BFO_struct}(a) Rhombohedral ($R3c$) unit cell of BiFeO$_3$. 
	 Because of the antiferromagnetic structure \new{and octahedral tilt pattern}, the unit cell contains two formula units. 
	 The ferroelectric polarization is directed along [111]. The Fe atoms are surrounded by tilted oxygen octahedra.
	 (b) Schematic band diagram for bulk BiFeO$_3$, with valence band maximum (VBM), conduction band minimum (CBM), 
electron polaron level $\varepsilon_p$,
   and energy barrier \new{$E_a$} for small electron polaron hopping between neighboring sites. 
   \new{$\langle\psi_i|H_c|\psi_f\rangle$ is the coupling matrix element between initial and final site of the hop, see section~\ref{sec:methods}.}
	 The computed barrier for hopping ($E_a\approx$\,0.2~eV) is considerably lower than the energy needed to transfer 
	 the electron from the polaron level $\varepsilon_p$ to the CBM ($\approx 0.7$ eV), see section \ref{sec:results}.
	 }
 \end{figure}

%%%%%%%%%%%%%%%%%%%%%%%%%%%%%%%%%%%%%%%%%%%%%%%%%%%%%
BiFeO$_3$ may be considered as a ferroelectric version of hematite, Fe$_2$O$_3$.
BiFeO$_3$ is a perovskite with a rhombohedral crystal structure at room temperature,
see \textbf{Figure}~\ref{fig:BFO_struct}(a),
a ferroelectric polarization of $\approx$100\,$\mu$C/cm$^2$ \cite{lebeugle:2007:very}, a bandgap 
of $\approx$2.7--3.0~eV \cite{basu:2008:photoconductivity,hauser:2008:characterization,ihlefeld:2008:optical,kumar:2008:linear,zelezny:2010:optical,sando:2018:revisiting,moubah:2012:photoluminescence}, 
 and a magnetic structure close to a collinear G-type antiferromagnet. 
 
Like hematite, BiFeO$_3$ might be suitable for water splitting. 
Other than hematite, BiFeO$_3$ \new{can form} ferroelectric domain walls, 
where electronic properties differ from those inside the bulk. % \cite{seidel:2009:conduction,bhatnagar:2013:role,guyonnet:2011:conduction,rojac:2017:domain,farokhipoor:2011:conduction}..
In particular, electrical conductivity has been observed at ferroelectric domain walls \new{in BiFeO$_3$ and in other ferroelectric perovskite oxides,}
whereas the surrounding bulk material is insulating or semiconducting \cite{seidel:2009:conduction,bhatnagar:2013:role,guyonnet:2011:conduction,rojac:2017:domain,farokhipoor:2011:conduction,risch:2022:giant}.
Like in \new{the case of} hematite \cite{lohaus:2018:limitation,peng:2012:semiconducting}, 
there is computational evidence that excess electrons in BiFeO$_3$ form small polarons with energy levels deep inside the band gap \cite{geneste:2019:polarons,radmilovic:2020:combined}, 
and that the small electron polarons are trapped by ferroelectric domain walls \cite{koerbel:2018:electron,koerbel:2020:photovoltage,koerbel:2023:optical}.
Hence, electronic transport in $n$-doped BiFeO$_3$ \new{should} occur by hopping of small electron polarons rather than by band-like transport, 
\new{if hopping of electron polarons requires less energy than promoting an electron from the polaron state to the conduction band,}
%see \textbf{Figure}~\ref{fig:scheme_energy_levels}.
see Figure~\ref{fig:BFO_struct}(b).
\new{Ferroelectric domain walls,} if present, could affect electronic transport.

\new{Here, for the first time, first-principles electronic structure calculations based on density-functional theory were performed to 
determine the energy barriers for electron polaron hopping in BiFeO$_3$. 
Different hopping directions were considered so that the anisotropy of the crystal structure was fully taken into account.
For the first time, polaron hopping at a ferroelectric domain wall was modeled from first principles. 
The neutral 71\textdegree~domain wall was chosen because 
it is one of the prevalent ferroelectric domain walls in BiFeO$_3$ \cite{dieguez:2013:domain,wang:2013:bifeo3},
because} it affects the atomic and electronic structure more strongly 
than the 109\textdegree~wall \cite{koerbel:2020:photovoltage,dieguez:2013:domain}, 
and because it has been reported to be conductive \new{(about one order of magnitude more than the domain interior)} in several studies, such as Ref.~\cite{farokhipoor:2011:conduction,chiu:2011:atomic,zhang:2019:intrinsic}.
The object of this study is to shed light on the intrinsic $n$-type conduction mechanism in BiFeO$_3$ 
and on the \new{influence} of pristine neutral ferroelectric domain walls.
%Both in bulk and in the domain wall, the obtained energy barriers are of the order of the thermal energy at room temperature,
%except for the energy barrier for escaping from a trapped state in the domain wall, which is about twice as large. 
%\begin{figure}[htb]
%	\includegraphics[width=0.3\textwidth]{Scheme_energy_levels.pdf}
%	\caption{\label{fig:scheme_energy_levels}Schematic band diagram for BiFeO$_3$ bulk, with valence band maximum (VBM), conduction band minimum (CBM), 
%	electron polaron level $\varepsilon_p$, 
%	and energy barrier for small electron polaron hopping between neighboring sites. 
%	The barrier for hopping ($\approx k_BT$ at room temperature) is lower than the energy needed to transfer the electron from the polaron level to the CBM ($\approx 0.7$ eV).}
%\end{figure}

This paper is organized as follows:
First the computational methods are introduced, 
then the \new{energy landscapes and hopping barriers for small polarons are presented for bulk and domain wall,}
and the electronic mobility and the diffusion constant at room temperature are determined.
Finally results are discussed based on experimental data, where available.
%%%%%%%%%%%%%%%%%%%%%%%%%%%%%%%%%%%%%%%%%%%%%%%%%%%%%
\section{\label{sec:methods}Methods}
%%%%%%%%%%%%%%%%%%%%%%%%%%%%%%%%%%%%%%%%%%%%%%%%%%%%%

The calculations were performed with the \new{``Vienna Ab initio Simulation Package'' ({\sc{Vasp}})} \cite{kresse:1996:efficiency},
using the Projector-Augmented Wave (PAW) method \cite {bloechl:1994:projector,kresse:1999:from} 
and pseudopotentials with 5 (Bi), 16 (Fe), and 6 (O) valence electrons, respectively.
Periodic boundary conditions were employed.
The local spin-density approximation (LSDA) to density-functional theory (DFT) was used, 
and the band gap was corrected with a Hubbard-$U$ of 5.3~eV applied to the Fe-$d$ states using Dudarev's scheme \cite{dudarev:1998:electron}.
This approach yields atomic and electronic structure in close agreement with experiment 
\cite{koerbel:2018:electron,koerbel:2020:photovoltage,koerbel:2023:optical}, \new{see also \textbf{Table}~\ref{tab:compare_xc_functionals}}.

Standard approaches to model polarons and polaron hopping from first principles are DFT+$U$ or hybrid DFT functionals, 
which were applied, e.g., in Ref.~\cite{setvin:2014:direct,ghorbani:2022:self}.
Here, to test how much the calculated hopping barriers depend on the level of theory,
the electronic structure with polaron and the polaron hopping barriers in bulk BiFeO$_3$ 
\new{from} LSDA+$U$ \new{are compared to those from the hybrid HSE06 functional \cite{heyd:03} \new{(25\% Hartree-Fock exchange, 5 \AA~screening length)} for the geometries optimized with LSDA+$U$.}
A similar comparison of the hybrid PBE0 functional \cite{adamo:1999:toward} and DFT+$U$ and a piece-wise linear DFT functional 
was made in Ref.~\cite{falletta:2023:polaron} for polarons in Ga$_2$O$_3$, 
with the result that polaron formation energy and eigenvalue from DFT+$U$ and PBE0 agree at least semiquantitatively. 

%This $U$ value was taken from the materials project \cite{jain:2013:materials}; 
%it was optimized with respect to oxide formation energies, but also yields band gaps and ferroelectric polarization close to experiment \cite{koerbel:2020:photovoltage}.
% With $U$=5.3~eV a ferroelectric polarization of $\approx 94\,\mu$C/cm$^2$ \cite{koerbel:2020:photovoltage} is obtained (Expt.: $\approx 100\,\mu$C/cm$^2$ \cite{lebeugle:2007:very}).
%%The reciprocal space was sampled with a $k$-point mesh of $8\times8\times8$ points in the first Brillouin zone of the 10-atom unit cell of the $R3c$ phase. 
Plane-wave basis functions with energies up to 520~eV were used.
Both the atomic positions and the cell parameters were optimized until 
the total energy differences between consecutive iteration steps 
fell below 0.01~meV for the optimization of the electronic density and below 0.1~meV for the optimization of the atomic structure.
%BiFeO$_3$ with excess electrons was modeled using supercells that comprise eight (bulk) or 16 (domain wall) 10-atom unit cells, respectively.

%%%%%%%%%%%%%%%%%%%%%%%%%%%%%%%%%%%%%%%%%%%%%%%%%%%%%
\subsection{Excess electrons in bulk}
%%%%%%%%%%%%%%%%%%%%%%%%%%%%%%%%%%%%%%%%%%%%%%%%%%%%%
BiFeO$_3$ bulk with excess electrons was modeled using supercells with 80 atoms
that consist of $2\times 2\times 2$ rhombohedral 10-atom unit cells.
\new{Convergence of energy barriers with respect to the supercell size was confirmed by comparing with results obtained with a 40-atom supercell, see Supp. Inf.
%Convergence of energy barriers with respect to supercell size was \new{investigated} for bulk, see Supporting Information.
The barriers obtained with the two supercell sizes differ only by about 10\%.}
The first Brillouin zone of the supercell was sampled with $2\times 2\times 2$ $k$-points, 
equivalent to $4\times 4\times 4$ $k$-points for the primitive rhombohedral ten-atom unit cell 
of BiFeO$_3$ and to $5\times 5\times 5$ $k$-points for a pseudocubic five-atom perovskite cell with about four \AA~edge length,
as depicted in \textbf{Figure}~\ref{fig:hopping_paths}(a).
\new{Convergence with respect to $k$-point density was confirmed by comparing with results obtained with a $4\times 4\times 4$ mesh, see Supp. Inf.}
%
%Systems without any excess charge carriers were considered as well as those
%with excess electrons or holes.
%The structures with excess charges were optimized in order to take into account polaron formation.
%
One excess electron was added to the supercell, whose charge was compensated by a uniform background charge density. 
The magnetic structure of BiFeO$_3$ was approximated by that of a collinear $G$-type antiferromagnet, spin-orbit coupling was neglected.

%%%%%%%%%%%%%%%%%%%%%%%%%%%%%%%%%%%%%%%%%%%%%%%%%%%%%
\subsection{Excess electrons at the ferroelectric \new{71\textdegree, 109\textdegree, and 180\textdegree}~domain wall\label{sec:methods:e_at_DW}}
%%%%%%%%%%%%%%%%%%%%%%%%%%%%%%%%%%%%%%%%%%%%%%%%%%%%%
The atomic structures and formation energies \new{(here  172\,mJ/m$^2$ for the 71\textdegree\ wall, 63 mJ/m$^2$ for the 109\textdegree\ wall, and 86 mJ/m$^2$ for the 180\textdegree\ wall \cite{koerbel:2020:photovoltage})} 
of low-energy ferroelectric domain walls in BiFeO$_3$ 
are well known \cite{dieguez:2013:domain,wang:2013:bifeo3,fousek:1969:orientation,ren:2013:ferroelectric,chen:2017:polar}. 
Atomic and electronic structure of the neutral 71\textdegree, \new{109\textdegree, and 180\textdegree}~domain walls investigated here 
have been published elsewhere \cite{koerbel:2018:electron, koerbel:2020:photovoltage,koerbel:2023:optical,dieguez:2013:domain,wang:2013:bifeo3}. 
\new{The computed properties of the ferroelectric 71\textdegree, 109\textdegree, and 180\textdegree\ domain walls, such as domain-wall energies and widths, 
can be found in the Supplemental Material to \cite{koerbel:2020:photovoltage}.}
The antiferromagnetic $G$-type spin configuration of the bulk was maintained in the systems with domain walls.
%%%%%%%%%%%%%%%%%%%%%%%%%%%%%%%%%%%%%%%%%%%%%%%%%%%%%
%\paragraph{Trapping energy}
%%%%%%%%%%%%%%%%%%%%%%%%%%%%%%%%%%%%%%%%%%%%%%%%%%%%%
\new{The energy profile of small electron polarons as a function of the distance from the domain walls was determined and the trapping energy at the domain walls was obtained.
240-atom supercells were employed with 12 perovskite layers in the direction perpendicular to the walls (six layers per domain, see \textbf{Figure}~\ref{fig:energy_dens_pol}).
$k$-point meshes of $1\times 2\times 2$ $k$-points were used. 
Very good convergence of results was obtained with this mesh, compared to a $1\times 4\times 4$ $k$-point mesh, see Supporting Information.
%Different $k$-point meshes ($1\times 2\times 2$ to $1\times 4\times 4$) were considered to ensure convergence; a $1\times 2\times 2$ mesh yields converged trapping energies, 
%see  Supporting Information.
The energy convergence thresholds were set to $10^{-6}$~eV for the electronic structure and to $10^{-5}$~eV for the atomic coordinates, respectively, which is probably one order of magnitude tighter than necessary.
The trapping energy $E_{\mathrm{trap}}$ was determined as the total energy difference between a polaron position in the domain interior (DI) 
and in the lowest-energy configuration at the domain wall (DW), 
$E_{\mathrm{trap}}=E(\mathrm{DW})-E(\mathrm{DI})$. 
Energy profiles and trapping energies were corrected by a factor $\Delta E_{\mathrm{FE,\ PBEsol}}/\Delta E_{\mathrm{FE,\ LSDA+}U}$, see~Table~\ref{tab:compare_xc_functionals}, 
to avoid underestimation of energy differences that originate in ferroelectricity.
Based on the computed trapping energies, the concentration $n$ of polarons was estimated based on a Boltzmann distribution, 
\begin{equation}
	n(s)=n_0\cdot \mathrm{e}^{-\frac{\Delta E(s)}{k_B\,T}},   
	\label{eq:conc}
\end{equation}
where $n_0$ is the concentration far away from the domain wall, $\Delta E(s)$ is the total energy difference between a polaron in $s$ and in the domain interior, 
$k_B$ is Boltzmann's constant, and $T$ is the temperature.
}

%%%%%%%%%%%%%%%%%%%%%%%%%%%%%%%%%%%%%%%%%%%%%%%%%%%%%
%\paragraph{Small polaron hopping}
%%%%%%%%%%%%%%%%%%%%%%%%%%%%%%%%%%%%%%%%%%%%%%%%%%%%%

\new{Hopping of small electron polarons at 71\textdegree~domain walls was} modeled using a 160-atom supercell, see Figure~\ref{fig:hopping_paths}(b),
spanned by  
4($\vec{a}_{\mathrm{pc}}+\vec{b}_{\mathrm{pc}}$) ($[110]$ in the pseudocubic system, here ``$s$''),
(-$\vec{a}_{\mathrm{pc}}+\vec{b}_{\mathrm{pc}}+2\vec{c}_{\mathrm{pc}}$ ($[\bar{1}12]$ in the pseudocubic system), 
and ($\vec{a}_{\mathrm{pc}}-\vec{b}_{\mathrm{pc}}+2\vec{c}_{\mathrm{pc}}$) ($[1\bar{1}2]$ in the pseudocubic system), 
where $\vec{a}_{\mathrm{pc}}$, $\vec{b}_{\mathrm{pc}}$, and $\vec{c}_{\mathrm{pc}}$ span the pseudocubic 5-atom cell. 
This supercell is based on that used in Ref.~\cite{koerbel:2018:electron,koerbel:2020:photovoltage,koerbel:2023:optical}.
\new{The first Brillouin zone of the supercell was sampled with $1\times 3\times 3$ $k$-points, 
equivalent to $5\times 5\times 5$ $k$-points for the 
primitive rhombohedral ten-atom unit cell 
of BiFeO$_3$ and to $7\times 7\times 7$ $k$-points for a 
pseudocubic five-atom perovskite cell with about four \AA~edge length.
}

%%%%%%%%%%%%%%%%%%%%%%%%%%%%
\subsection{Polaron hopping}
%%%%%%%%%%%%%%%%%%%%%%%%%%%%
According to previous works \cite{geneste:2019:polarons,radmilovic:2020:combined,koerbel:2018:electron,koerbel:2020:photovoltage,koerbel:2023:optical}, 
the excess electron localizes mostly on a single Fe site.
In bulk BiFeO$_3$, all Fe sites are equivalent. 
Localization on a specific Fe site was induced by means of the occupation matrix control (OMC) method of Allen and Watson~\cite{allen:2014:occupation}.
%Without OMC, a small polaron forms spontaneously on one of the Fe sites, if symmetry allows it.

Energy barriers for polaron hopping between %second-nearest neighbor 
different Fe sites were obtained using the nudged-elastic-band (NEB) method, 
in which an artificial spring force acting along the hopping path constrains atomic structure optimization to directions 
perpendicular to the \new{hopping} path. Nine configurations (NEB images) along each hopping path were considered, including initial and final state.
\new{The NEB method is designed to determine the minimum-energy path, 
which can have a considerably lower energy barrier than the shortest path (the linear path). %, the linear interpolation between initial and final configuration, see Supporting Information.}
Since here the top of the barrier is pointed (the forces along the minimum energy path are not continuous), 
which might pose numerical difficulties for the NEB algorithm,
the top of the barrier was obtained by calculating the energy along a linear path between the two nearest NEB images (images 4 and 5).
The deviation of the barrier from the highest NEB energies obtained for the 2NN hops in bulk were then used to estimate the barrier tops for hops at the domain wall, 
i.e. the energy of the linear path (between images 4 and 5) was not calculated for the domain-wall hops.
}
Both hops between nearest-neighbor (NN) and second-nearest neighbor (2NN) Fe sites were considered, 
but NN hops require spin flip of the electron and should hence be suppressed.

%Mainly hopping between second-nearest neighbor sites was considered, since nearest-neighbor hopping would require a spin flip of the electron.

\begin{figure*}[htb]
 \includegraphics[width=0.35\textwidth]{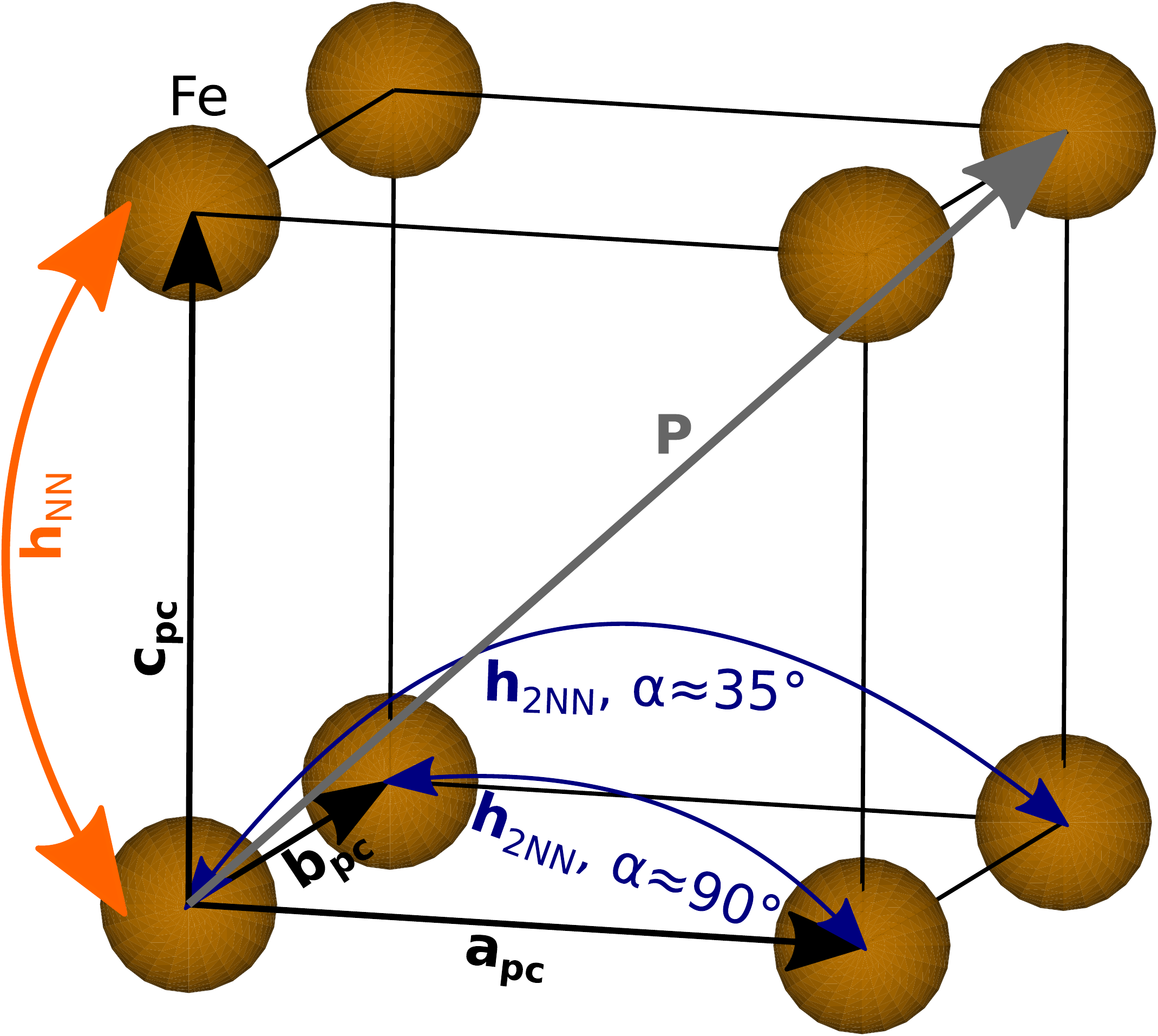}
 \includegraphics[width=0.64\textwidth]{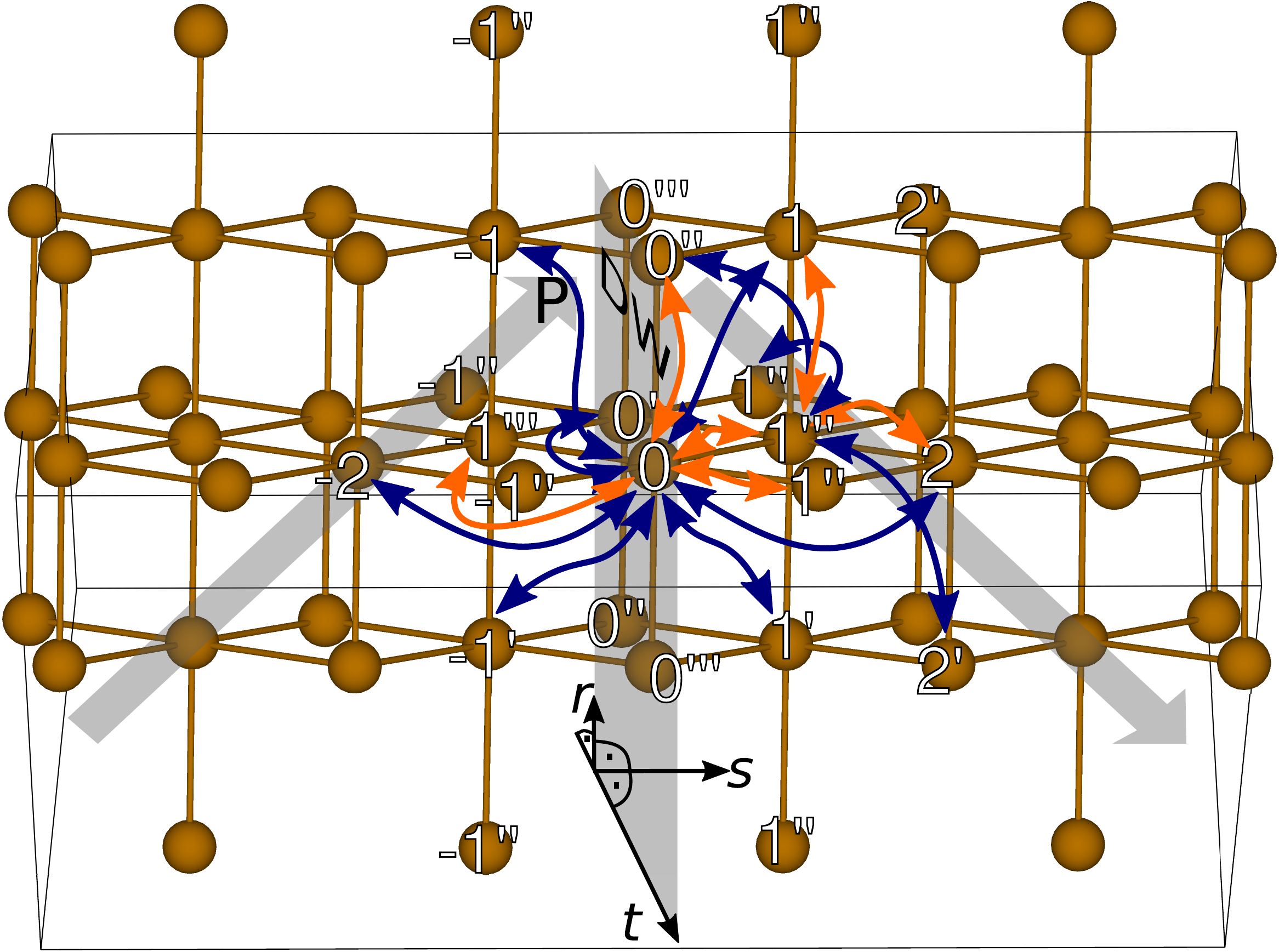}
 \put(-500,240){(a)}
 \put(-340,240){(b)}
	\caption{\label{fig:hopping_paths}(a) Fe network (schematic) and hops $\vec{h}$ between nearest neighbors (NN, orange) and second-nearest neighbors (2NN, dark-blue) in bulk BiFeO$_3$. 
	For clarity, only Fe atoms are shown.
	There is one irreducible hop between nearest neighbors and
	two inequivalent hops between second-nearest neighbors that form angles 
	with the ferroelectric polarization $\vec{P}$ of $\alpha\approx$\new{90}\textdegree~and $\alpha\approx\,35$\textdegree, re\-spec\-tive\-ly.
	(b) 
	160-atom supercell with the 71\textdegree~domain wall (DW) and hopping paths considered. % (orange: NN, dark-blue: 2NN hops). 
	%The vectors $\vec{e}_s$, $\vec{e}_r$, and $\vec{e}_{t}$ are an orthonormal set. $\vec{e}_r$ and $\vec{e}_t$  span the domain-wall plane, $\vec{e}_s$ is perpendicular to the domain wall, 
	%$\vec{e}_{t}$ is perpendicular to the polarization $\vec{P}$. 
	The coordinates $s$, $r$, and $t$ are orthogonal. $r$ and $t$ span the domain-wall plane, $s$ is perpendicular to the domain wall, 
	$t$ is perpendicular to the polarization $\vec{P}$. 
	 }
\end{figure*}

The number of inequivalent NN and 2NN hops is limited by symmetry.
In bulk, there \new{are} one NN hop and 
two inequivalent 2NN hops: a 2NN hop perpendicular to the ferroelectric polarization direction, 
and a 2NN hop in a direction that forms an angle of $\alpha\approx\,35$\textdegree~with the polarization direction,
see \textbf{Figure}~\ref{fig:hopping_paths}(a).
At ferroelectric domain walls, Fe sites are still equivalent within planes parallel to the wall, but differ between these planes. 
Hops can occur along the wall or with a component perpendicular to the wall, see Figure~\ref{fig:hopping_paths}(b). %\ref{fig:DW_geometry}.
\new{The} NEB calculations were considered converged when the energy barrier changed by maximally 1~meV in two consecutive iterations of the NEB structure optimization.
The strain dependence of the barriers was also investigated.
The strain reference (zero strain, $\varepsilon=0$) corresponds to the optimized geometry of bulk or domain wall without excess electron.
Tensile strain, $\varepsilon>0$, corresponds to the optimized geometry of bulk or domain wall with an excess electron.

%\begin{figure*}[htb]
% %\includegraphics[width=0.45\textwidth]{CONTCAR_DW.pdf}
% %\def\svgwidth{0.4\textwidth}
% %\input{CONTCAR_latex.pdf_tex}
% %\includegraphics[width=0.45\textwidth]{CONTCAR_vesta_pdf.pdf}
% \includegraphics[width=0.7\textwidth]{CONTCAR_cleaned.pdf}
%	\caption{\label{fig:DW_geometry}160-atom supercell with the 71\textdegree~domain wall (DW) and hopping paths considered (orange: NN, dark-blue: 2NN hops). 
%	For clarity, only Fe atoms are shown.
%	%The vectors $\vec{e}_s$, $\vec{e}_r$, and $\vec{e}_{t}$ are an orthonormal set. $\vec{e}_r$ and $\vec{e}_t$  span the domain-wall plane, $\vec{e}_s$ is perpendicular to the domain wall, 
%	%$\vec{e}_{t}$ is perpendicular to the polarization $\vec{P}$. 
%	The coordinates $s$, $r$, and $t$ are orthogonal. $r$ and $t$ span the domain-wall plane, $s$ is perpendicular to the domain wall, 
%	$t$ is perpendicular to the polarization $\vec{P}$. 
%	 }
%\end{figure*}
%

%%%%%%%%%%%%%%%%%%%%%%%%%%%%%%%%%
\subsection{Transition state theory}
%%%%%%%%%%%%%%%%%%%%%%%%%%%%%%%%%
The abrupt jump of the small electron polaron from initial to final site (see below) as a function of the hopping coordinate 
indicates that the hopping process is non-adiabatic rather than adiabatic.
Marcus theory \cite{marcus:1985:electron} was applied to obtain non-adiabatic hopping rates of small electron polarons between 2NN lattice sites in BiFeO$_3$.
%The general approach is well-known and was applied, e.g., to ...\cite{}. %small electron polaron hopping in hematite assuming adiaba\cite{Neaton}.

NN hopping, being spin-forbidden, was neglected.  
The transition rate $k$ is then
\begin{equation}
	\label{eq:Marcus}
  %k=\frac{2\pi}{\hbar}|\langle\psi_i|H_c|\psi_f\rangle|^2\frac{1}{\sqrt{4\pi\lambda k_BT}}\mathrm{exp}\left(-\frac{(\lambda+\Delta G^0)^2}{4\lambda k_BT}\right).
  k=\frac{2\pi}{\hbar}|\langle\psi_i|H_c|\psi_f\rangle|^2\frac{1}{\sqrt{4\pi\lambda k_BT}}\mathrm{exp}\left(-\frac{E_a}{k_BT}\right).
\end{equation}
$E_a$ is the activation energy (hopping barrier) for the 2NN hop, $\lambda$ is the reorganization energy.
For symmetric hops, $\lambda=4 E_a$.
The electronic coupling matrix elements $\langle\psi_i|H_c|\psi_f\rangle$, 
where the $\psi_{i,f}$ are electronic wave functions localized on the initial and final Fe site, respectively, 
were adopted from Ref.~\cite{newton:1980:formalisms}, 
where they were calculated for Fe(H$_2$O)$_6^{5+}$ complexes as a function of the Fe-Fe distance from a model Hamiltonian 
that treated Fe with a pseudopotential and H$_2$O as classical point charges. 
\new{Rosso and coworkers obtain a coupling matrix element of 60\,meV for second-nearest neighbors in hematite \cite{rosso:2003:ab}, 
where the intersite distance is smaller and electronic intersite coupling should be larger than in BiFeO$_3$.
}
Here, the matrix element for the second-nearest Fe-Fe distance calculated for bulk BiFeO$_3$, \new{5.516}\,$\AA$, was used throughout (-101.875\,cm$^{-1}\approx -0.013$\,eV). % -0.0126 eV, from the literature
Alternatively, the coupling matrix elements could \new{in principle} be determined as the deviation of the calculated barrier from the parabolic crossing point,
%see \textbf{Figure}~\ref{fig:scheme_VAB_and_DOS}(a). 
see Figure~\ref{fig:BFO_struct}(b). %\new{and \ref{fig:energy_NEB_bulk_hires}.}
%In the case of compressive strain, where the calculated hopping energy profile is close to a double parabola, 
%one obtains a coupling matrix element of $\langle\psi_i|H_c|\psi_f\rangle\approx -2$ to $-3$\,meV.
\new{In this case one obtains a coupling matrix element $\langle\psi_i|H_c|\psi_f\rangle$ of the order of $-1$\,meV, see Supp. Inf., 
with a large error bar. This result is not used here.
}

\begin{figure}
	\includegraphics[width=0.65\textwidth]{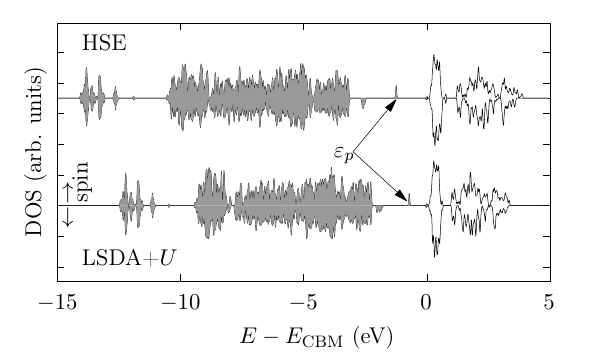}
	\put(-510,190){(a)}
	\put(-350,190){(b)}
	\caption{\label{fig:scheme_VAB_and_DOS}(a) %Extracting the coupling matrix element $\langle \psi_i|H_c|\psi_f\rangle$ between initial state (i) and final state (f) from the hopping energy profile.
	Electronic density of states \new{of bulk BiFeO$_3$ with an} electron polaron from LSDA+$U$ and HSE06.
	The energy of the conduction band minimum (CBM) is set to zero. 
	$\varepsilon_p$ is the computed polaron level. 
       %(b)
	}
\end{figure}

The electron mobility $\mu_e$ and the diffusion constant $D$ are obtained from the transition rate $k$ using the Einstein relation as
%\begin{eqnarray}
%	\mu_e&=&\frac{e\, |\vec{h}|^2\, n_f\,k}{2\,d\,k_B\,T}=\frac{D\, e}{k_B\,T},\\
%	D&=&\frac{|\vec{h}|^2\, n_f\,k}{2\,d},\nonumber
%\end{eqnarray}
\begin{equation}
	\label{eq:mu_bulk}
	\mu_e=\frac{D\, e}{k_B\,T}=\frac{e\, |\vec{h}|^2\, n_f\,k}{2\,d\,k_B\,T},
\end{equation}
where $e$ is the elementary charge, $|\vec{h}|$ is the hopping distance,
$n_f$ is the number of equivalent available final states for a given initial state, 
$d$ is the dimension,
$k_B$ is Boltzmann's constant, and $T$ is the temperature, here 300\,K.
For bulk, $d$=3, $n_f^{\mathrm{bulk}}$=12 for 2NN hops,
$|\vec{h}|=\sqrt{2}\,a_0$ is the 2NN Fe-Fe distance, and $a_0$ the pseudocubic lattice constant (about 4\,\AA).
For two-dimensional transport within a domain-wall plane,
$d=2$. Hops along the $r$ coordinate consist of two consecutive hops,
\new{such as $0'''\to 1'''\to 0''$ or $0'''\to -1'''\to 0''$, with $|\vec{h_r}|=a_0$ and $n_{f,r}=8$. 
Hops along the $t$ coordinate can be single 2NN hops with $|\vec{h_t}|=\sqrt{2}a_0$ and $n_{f,t1}$=2,
such as $0\to 0'$, or double 2NN hops with $n_{f,t2}=8$ and $|\vec{h_t}|=a_0/\sqrt{2}$, such as $0'\to 1\to 0$ or $0'\to -1\to 0$.}
%The effective barrier for double hops was taken as the higher one  of the two.
The two-dimensional $n$-type mobility and the diffusion constant in the domain wall are computed as
%\begin{eqnarray}\label{eq:mu_DW}
%	\mu_e^{\mathrm{DW}}&=&\frac{e}{4\,k_B\,T} \left( k_r^{\mathrm{eff}} |\vec{h}_r|^2\, n_{f,r} +  k_{t1}^{\mathrm{eff}} |\vec{h}_t|^2\, n_{f,t1} +  k_{t2}^{\mathrm{eff}} |\vec{h}_t|^2\, n_{f,t2} \right),\\\nonumber
%	D^{\mathrm{DW}}&=&k_B\,T\,\mu_e^{\mathrm{DW}}/e
%\end{eqnarray}

\begin{equation}\label{eq:mu_DW}
	\mu_e^{\mathrm{DW}}=\frac{e\,D^{\mathrm{DW}}}{k_B\,T}=\frac{e}{4\,k_B\,T} \left( k_r^{\mathrm{eff}} |\vec{h}_r|^2\, n_{f,r} +  k_{t1}^{\mathrm{eff}} |\vec{h}_t|^2\, n_{f,t1} +  k_{t2}^{\mathrm{eff}} |\vec{h}_t|^2\, n_{f,t2} \right)\new{=\mu_{e,r}^{\mathrm{DW}}+\mu_{e,t}^{\mathrm{DW}}}.
\end{equation}

%%%%%%%%%%%%%%%%%%%%%%%%%%%%%%%%%
\section{\label{sec:results}Results}
%%%%%%%%%%%%%%%%%%%%%%%%%%%%%%%%%
%%%%%%%%%%%%%%%%%%%%%
\subsection{\new{Material properties of BiFeO$_3$ obtained with different exchange-correlation functionals}}
%%%%%%%%%%%%%%%%%%%%%
\new{\textbf{Table}~\ref{tab:compare_xc_functionals} contains properties of 
BiFeO$_3$ in the cubic and the rhombohedral phase obtained with different exchange-correlation functionals $f_{xc}$ 
and from experiment. The calculations were performed using the primitive rhombohedral 10-atom unit cell of BiFeO$_3$.
A cutoff energy of 520 eV, a $k$-point mesh of $8\times 8\times 8$ points, and energy convergence thresholds of $10^{-8}$ eV and $10^{-7}$ eV for electronic and atomic structure were employed, respectively, 
except for HSE06, where a $k$-point mesh of $6\times 6\times 6$ points was used.
The ionic polarization $P=|\vec{P}|$ was computed as a weighted sum of atomic displacements, 
\begin{equation}
	\vec{P}=\frac{1}{V}\sum\limits_{i}\left(\vec{r}_{i}-\vec{r}_{i}^{(0)}\right)\cdot q_{i}\cdot \mathrm{w}_{i}.
	\label{eq:pol}
\end{equation}
Here, $i$ denotes atoms in the unit cell, $\vec{r}_i$ and $\vec{r}_i^{(0)}$ are actual positions in the ferroelectric (FE) and reference positions in the paraelectric (PE) phase, 
the charges $q_i$ are nominal ionic charges (Bi$^{3+}$, Fe$^{3+}$, O$^{2-}$), the $w_i$ are weighting factors, and $V$ is the unit-cell volume.
}

\begin{table}[htb]
 \begin{tabular}{l|l|l|l|l|l}
	 $f_{\mathrm{xc}}$ &    $\Delta E_{\mathrm{FE}}$ (eV) &  $a_c$ (\AA) & $a_r$ (\AA) &  $\alpha$ (\textdegree) & $P$ ($\mu$C/cm$^2$) \\\hline 
 HSE06 & -1.15 & 5.48 & 5.63 & 59.15 & 73 \\
LSDA & -0.55 & 5.26 & 5.43 & 60.53 & 63 \\
LSDA+$U$ & -0.90 & 5.44 & 5.52 & 59.82 & 61 \\
PBE & -0.73 & 5.54 & 5.59 & 59.44 & 52 \\
PBE+$U$ & -0.84 & 5.56 & 5.71 & 59.05 & 68 \\
PBEsol & -1.00 & 5.46 & 5.54 & 59.93 & 66 \\
expt. &  &  & 5.64 & 59.42 & 72 \\
 \end{tabular}
	\caption{\label{tab:compare_xc_functionals}\new{Ferroelectric energy gain $\Delta E_{\mathrm{FE}}=E_{\mathrm{total,\ FE}}-E_{\mathrm{total,\ PE}}$ per formula unit, cubic and rhombohedral lattice parameter $a_c$ and $a_r$, 
	rhombohedral lattice angle $\alpha$, and ionic polarization $P$ obtained with different exchange-correlation functionals $f_{xc}$ and from experiment \cite{moreau:1971:ferroelectric}}.}
\end{table}
\new{Taking PBEsol or HSE06, which yield excellent structural agreement with experiment, as a benchmark, LSDA+$U$ underestimates the ferroelectric energy gain by about 10--20\%.
}
%%%%%%%%%%%%%%%%%%%%%
\subsection{Excess electrons in bulk}
%%%%%%%%%%%%%%%%%%%%%
%%%%%%%%%%%%%%%%%%%%%%%%%%%%%%%%%%%%%%%%%%%%%%%%%%%%%%%%%%%
\threesubsection{Electronic structure from LSDA+$U$ and HSE}\\
%%%%%%%%%%%%%%%%%%%%%%%%%%%%%%%%%%%%%%%%%%%%%%%%%%%%%%%%%%%
\new{
\begin{table}
  \caption{\label{tab:pol_props_bulk}\new{Computed polaron formation energy $E_p^f$ and eigenvalue $\varepsilon_p$ with respect to the conduction band minimum
  of small electron polarons in bulk BiFeO$_3$ from LSDA+$U$ and from HSE06 for strain $\varepsilon$=0.
    }}
	\begin{tabular}{l | c | c }
	\hline
			                   & LSDA+$U$ & HSE06   \\\hline      
        $E_p^f$ (eV)                 &  -0.34 &  -0.57   \\
         $\varepsilon_p$  (eV)   &  -0.71 &  -1.2     
	\end{tabular}
\end{table}
}
\new{The formation energy $E_p^f$ of small polarons is strongly negative (LSDA$+U$: $E_p^f\approx-$0.3\,eV, HSE: $E_p^f\approx-$0.6\,eV, see \textbf{Table~\ref{tab:pol_props_bulk}}), 
    hence the small polaron is the ground state and should be stable not only at low temperature, but also \new{at room temperature and above}.
    }
The electronic density of states (DOS) for a supercell (80 atoms) with an electron polaron is shown in \textbf{Figure}~\ref{fig:scheme_VAB_and_DOS}. %(b). %\textbf{Figure}~\ref{fig:DOS}.}
%\begin{figure}
%	\includegraphics[width=0.5\textwidth]{DOS.pdf}
%	\caption{\label{fig:DOS}\new{Electronic density of states with electron polaron from LSDA+$U$ and HSE.
%	The energy of the conduction band minimum (CBM) is set to zero. 
%	$\varepsilon_p$ is the computed polaron level.
%	}
%	}
%\end{figure}
The DOS obtained with LSDA+$U$ and HSE06 agree semiquantitatively.
The computed polaron level $\varepsilon_p$ lies 0.7\,eV below the conduction band minimum for LSDA+$U$ and 1\,eV below the conduction band minimum for HSE06,
 with an estimated uncertainty of 0.2~eV. \new{Hence the polaron level is a deep level at room temperature.}
	
\begin{figure}[h]
	    \includegraphics[width=0.49\textwidth]{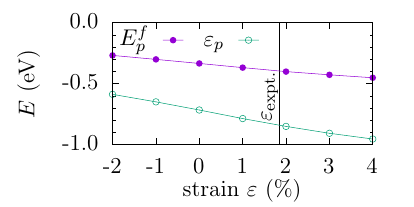}
	    \caption{\label{fig:E_pol_strain}\new{Polaron formation energy $E_p^f$ and polaron eigenvalue $\varepsilon_p$ as a function of strain from LSDA+$U$. 
	    Zero strain corresponds to the computationally optimized crystal structure. $\varepsilon_{\mathrm{expt.}}$ corresponds to the experimental volume at room temperature 
	    \cite{moreau:1971:ferroelectric}.}}
    \end{figure}

\new{The moderate strain dependence of polaron formation energy $E_{p}^f$ and electronic eigenvalue $\varepsilon_p$ is depicted in \textbf{Figure}~\ref{fig:E_pol_strain}.}
%
%%%%%%%%%%%%%%%%%%%%%%%%%%%%%%%%%%%%%%%%%%%%%%%%%%%%%%%%%%%%%%%%%%
%\subsubsection{Second-nearest neighbor hops, spin preserved}
%%%%%%%%%%%%%%%%%%%%%%%%%%%%%%%%%%%%%%%%%%%%%%%%%%%%%%%%%%%%%%%%%%
%%%%%%%%%%%%%%%%%%%%%%%%%%%%%%%%%%%%%%%%%%%%%%%%%%%%%%%%%%%
%\\
\threesubsection{Hopping barriers}\\
%%%%%%%%%%%%%%%%%%%%%%%%%%%%%%%%%%%%%%%%%%%%%%%%%%%%%%%%%%%
\begin{figure*}[h]
 \includegraphics[width=0.99\textwidth]{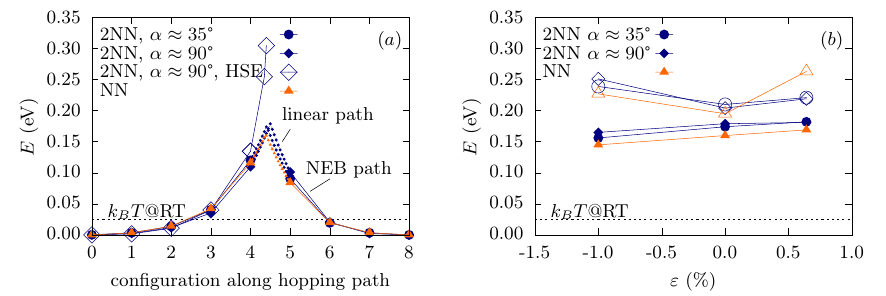}
 \includegraphics[width=0.24\textwidth]{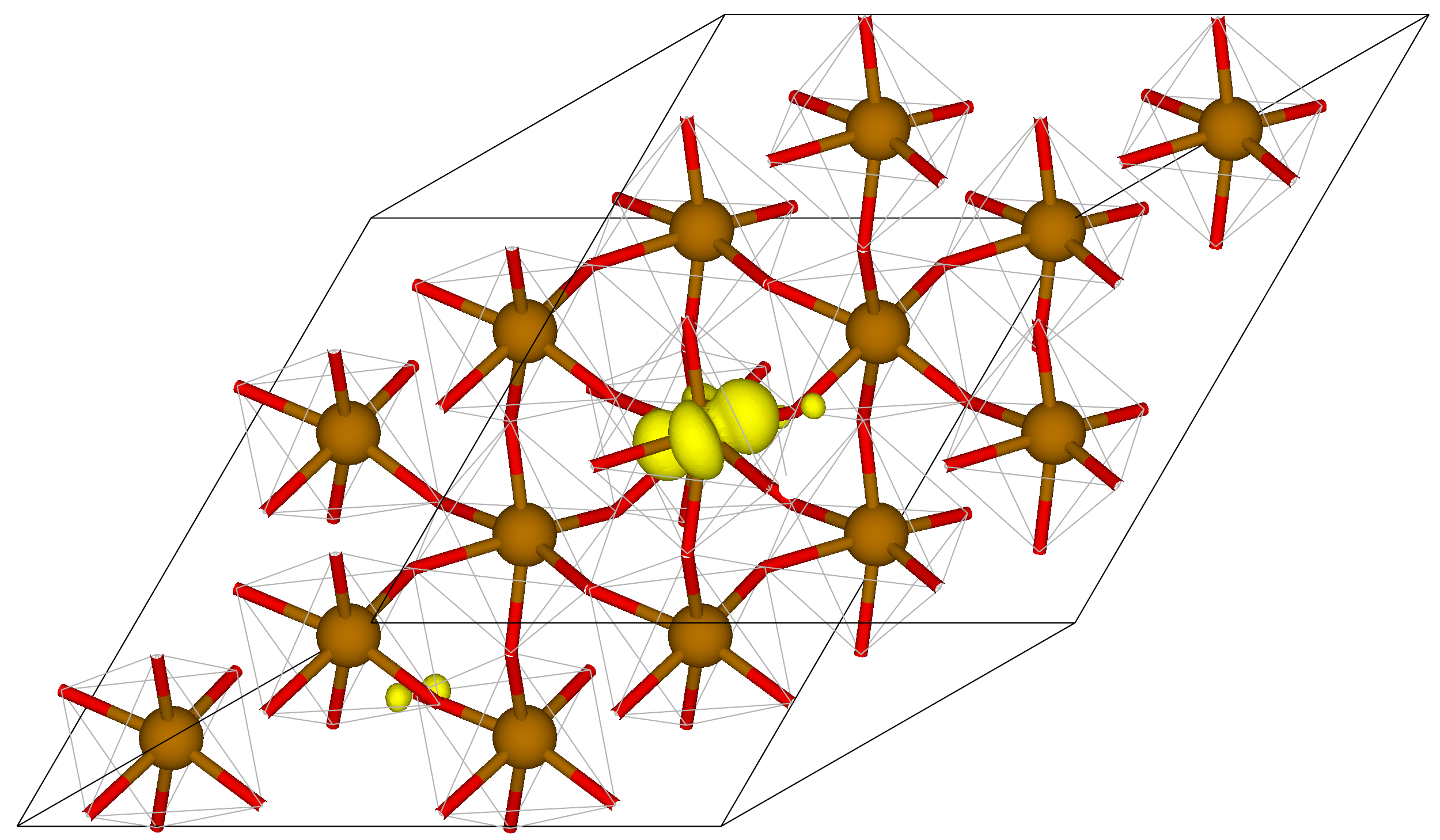}
 \includegraphics[width=0.24\textwidth]{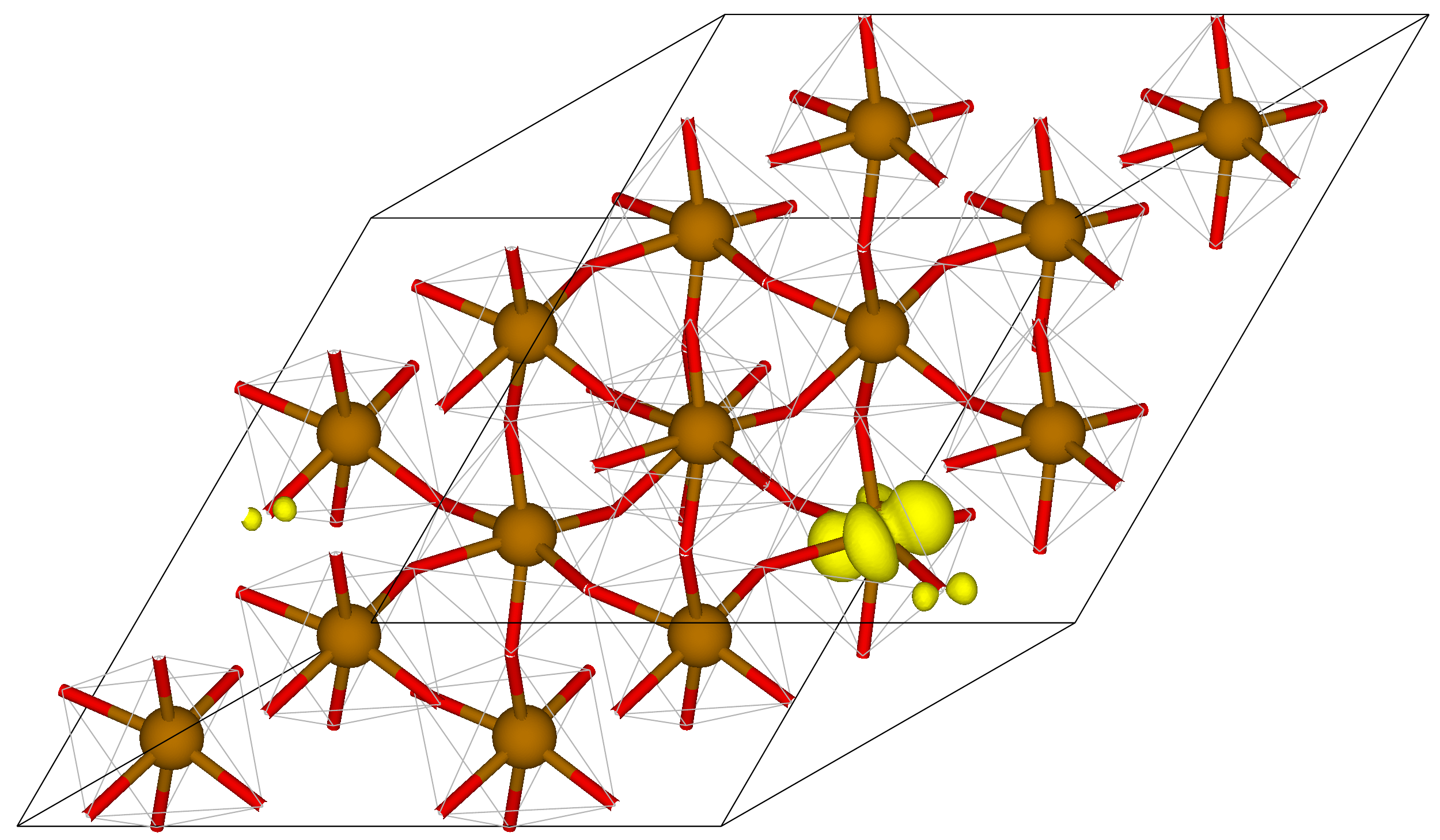}
 %\includegraphics[width=0.24\textwidth]{PARCHG_img8.pdf}
	%\put(-510,50){(c)}
	\put(-260,65){(c) $|\mathrm{initial}\rangle$}
	\put(-120,65){(d) $|\mathrm{final}\rangle$}
	%\put(-115,50){(f)}
	\caption{\label{fig:bulk_hops} (a) Energy profile of first- (orange triangles) and second-nearest neighbor hops (dark-blue circles and diamonds) in bulk for zero strain  
	along the hopping path. 
	The configurations \new{0 to 4 and 5 to 8 in (a)} belong to initial and final state, respectively. 
	(b) Hopping barriers of NN and 2NN hops as a function of strain $\varepsilon$.
	The lines are a guide to the eye. Solid symbols: LSDA+$U$, empty symbols: HSE06.
	(c) and (d): isosurface of the density of the excess electron 
	(yellow; isosurface level: $\approx$2\% of the maximum) of configuration \new{4 and 5} of the 2NN hop with $\varepsilon=0$ and $\alpha\approx \new{90}$\textdegree.
	Fe atoms are shown as brown spheres.
	%, the isosurface of the density of the excess electron (at $\approx$2\% of the maximum) along the 2NN hopping path is shown in yellow.
	}
\end{figure*}
The energy barriers for NN hops and the two different 2NN hops are depicted in \textbf{Figure}~\ref{fig:bulk_hops}(a) and (b) for different strains $\varepsilon$. 
Both NN and 2NN barriers are \new{about 0.2~eV and slightly increase under tensile strain}.
Hence, the hopping barrier is \new{considerably} smaller than the energy distance \new{of about 0.7\,eV} to the conduction band minimum.
The 2NN barriers are not strongly direction dependent. 
The transitions from initial to final site are non-adiabatic (abrupt), 
as seen from the isosurface of the excess electron's density as a function of the hopping coordinate 
(yellow orbitals in Figure~\ref{fig:bulk_hops}\new{(c) and (d)}). 
%The transitions between sites are non-adiabatic (abrupt jumps from initial to final site), see Fig.~\ref{fig:bulk_hops}, 
%which depicts the isosurface of the excess electron along the 2NN hopping path with $\alpha\approx 0$\textdegree\ in bulk.
%\begin{figure*}[htb]
%	\includegraphics[width=0.1\textwidth]{2NN_path_PARCHG_00.pdf}
%	\includegraphics[width=0.1\textwidth]{2NN_path_PARCHG_01.pdf}
%	\includegraphics[width=0.1\textwidth]{2NN_path_PARCHG_02.pdf}
%	\includegraphics[width=0.1\textwidth]{2NN_path_PARCHG_03.pdf}
%	\includegraphics[width=0.1\textwidth]{2NN_path_PARCHG_04.pdf}
%	\includegraphics[width=0.1\textwidth]{2NN_path_PARCHG_05.pdf}
%	\includegraphics[width=0.1\textwidth]{2NN_path_PARCHG_06.pdf}
%	\includegraphics[width=0.1\textwidth]{2NN_path_PARCHG_07.pdf}
%	\includegraphics[width=0.1\textwidth]{2NN_path_PARCHG_08.pdf}
%	\caption{\label{fig:hopping_parchg} Isosurface of the density of the excess electron along the 2NN hopping path (from left to right) with $\alpha\approx 0$\textdegree\ in bulk. 
%	For clarity, only Fe atoms are shown. 
%	}
%\end{figure*}
The computed energy barriers are listed in \textbf{Table}~\ref{tab:barriers_bulk}.
\begin{table}
	\caption{\label{tab:barriers_bulk}Computed energy barrier $E_a$\new{, electron mobility $\mu_e$, and diffusion constant $D$} for small electron polaron hopping in bulk BiFeO$_3$ 
	from LSDA+$U$ and from HSE06 for different strains. \new{1.9\%} strain correspond to the experimental cell volume at room temperature.
	The estimated uncertainty is $\approx$20\,meV.
	}
	\begin{tabular}{l | c | c | c| c | c | c | c | c }
	\hline
			       & \multicolumn{6}{c|}{$E_a$ (eV)}            &      &          \\
		$\varepsilon$  & \multicolumn{2}{c |}{NN}   & \multicolumn{2}{c |}{2NN, $\alpha\approx 90$\textdegree} & \multicolumn{2}{c |}{2NN, $\alpha\approx 35$\textdegree } & & \\
			       & LSDA+$U$ & HSE06  &  LSDA+$U$ & HSE06         & LSDA+$U$  & HSE06 &$\mu_e$     (cm$^2$/Vs)    & $D$  (cm$^2$/s)      \\\hline      
		$-1.0$\%       &  0.15    &  0.23  &   0.16 &  0.25    &  0.16      & 0.24 & -- & --\\
		 0             &  0.16    &  0.19  &   0.18 &  0.20    &  0.17      & 0.21 & -- & --     \\
		 0.64\%        &  0.17    &  0.26  &   0.18 &  0.22    &  0.18      & 0.22 & $0.3\dots 1 \cdot 10^{-3}$ & $0.8\dots 4 \cdot 10^{-5}$    \\	
		 1.9\%         &  0.19    &  --    &   0.18 &  --      &  0.20      &  -- & $0.2\dots 1 \cdot  10^{-3}$ & $0.5\dots 3 \cdot 10^{-5}$    \\	\hline
	\end{tabular}
\end{table}
\new{The barrier obtained with HSE06 is considerably larger, however the structure was not optimized with HSE06 but with LSDA+$U$, therefore the barrier from HSE06 should be overestimated. 
In the following, the barrier obtained with LSDA+$U$ is used.}
To obtain the electron mobility and the diffusion constant at room temperature, 
the extrapolated barrier at $\varepsilon\approx$1.9\%, corresponding to the experimental lattice constant at room temperature \cite{moreau:1971:ferroelectric}, 
was used. With an energy barrier of \new{0.19\,eV} (the average of the two 2NN barriers)
and a hopping distance of the 2NN Fe-Fe distance,
the computed electron mobility in bulk BiFeO$_3$ at room temperature is \new{$\mu_e^{\mathrm{bulk}}\approx 5\cdot 10^{-4}$\,cm$^2$/Vs,
the computed diffusion constant is $D \approx 1\cdot 10^{-5}$\,cm$^2$/s. With an estimated uncertainty of $\pm$10\% in the activation barrier,  
one obtains $\mu_e^{\mathrm{bulk}}\approx 0.2$ to $1\cdot 10^{-3}$\,cm$^2$/Vs and $D \approx 0.5$ to $3\cdot 10^{-5}$\,cm$^2$/s.
}

%%%%%%%%%%%%%%%%%%%%%%%%%%%%%%%%%%%%%%%%%%%%%%%%%%%%%%%%%%%%%%%%%%
%\subsubsection{Nearest neighbor hops, spin-flip needed}
%%%%%%%%%%%%%%%%%%%%%%%%%%%%%%%%%%%%%%%%%%%%%%%%%%%%%%%%%%%%%%%%%%
%

%%%%%%%%%%%%%%%%%%%%%
\subsection{Excess electrons at domain walls}
%%%%%%%%%%%%%%%%%%%%%
%%%%%%%%%%%%%%%%%%%%%
\subsubsection{\new{Trapping of polarons at domain walls}}
\begin{figure}[htb]
	\includegraphics[width=0.33\textwidth]{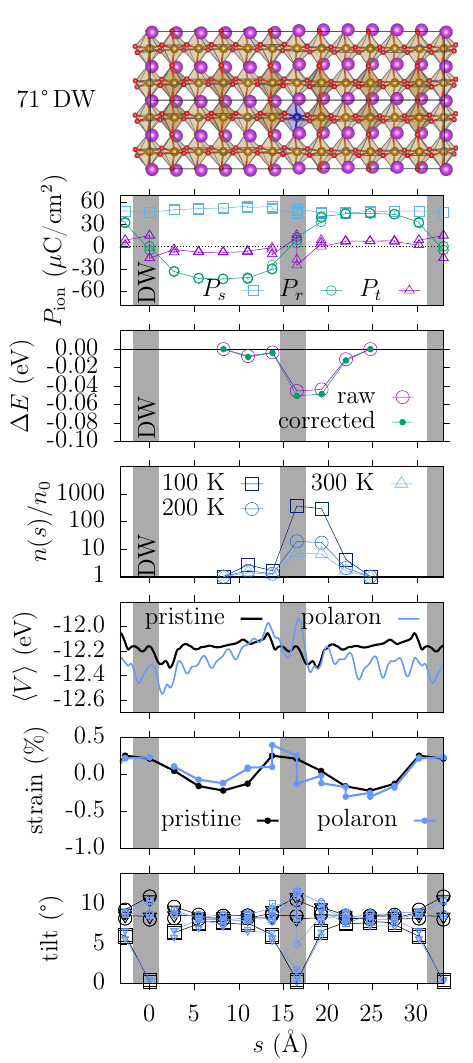}
	\includegraphics[width=0.33\textwidth]{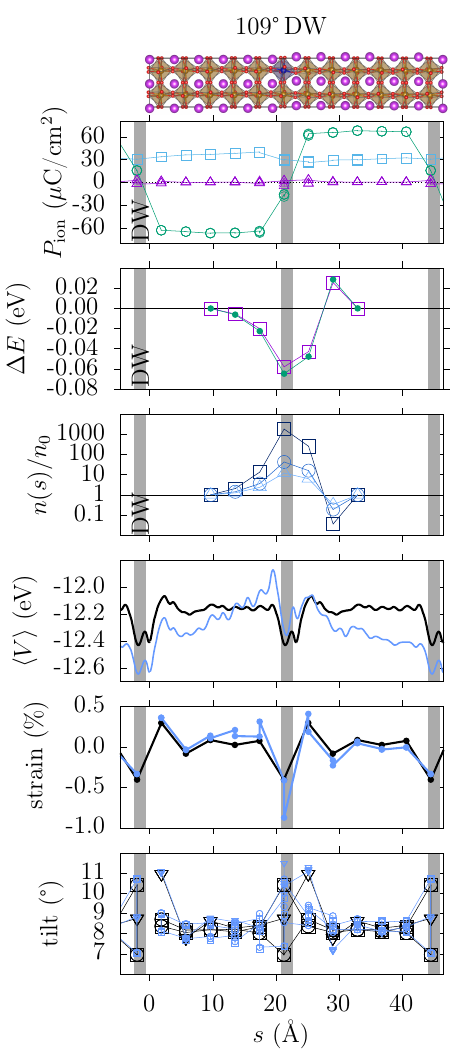}
	\includegraphics[width=0.33\textwidth]{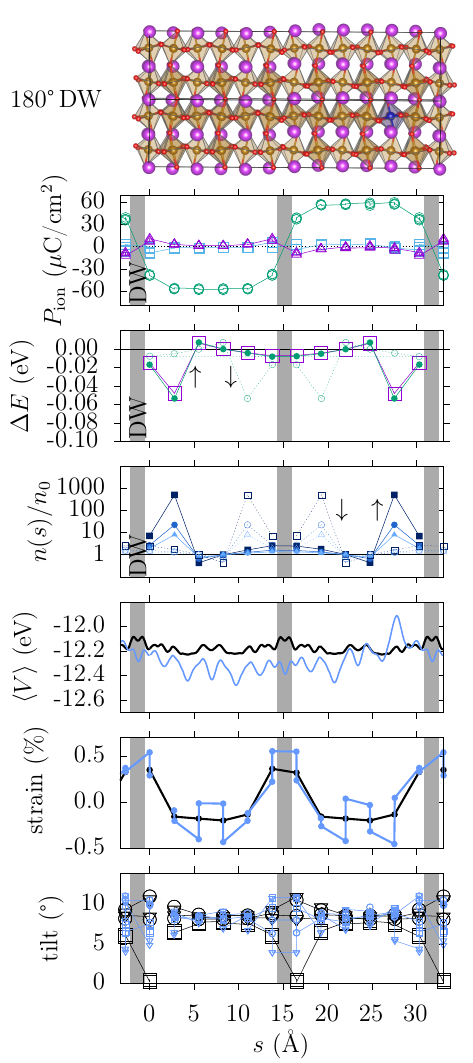}
	\caption{\label{fig:energy_dens_pol}\new{From top to bottom: atomic structure, ionic polarization profile, energy profile, density, electronic potential, local strain profile, and octahedral tilt, 
	for small electron polarons at the 71\textdegree\ (left), 109\textdegree\ (center), and the 180\textdegree\ DW (right).
	For the 180\textdegree\ wall, energy profiles and densities of spin-up ($\uparrow$) and spin-down ($\downarrow$) polarons are shown separately, since this wall is spin-selective.}
	}
\end{figure}

%%%%%%%%%%%%%%%%%%%%%
%
\new{
	The atomic structure, ionic ferroelectric polarization profile $P_{\mathrm{ion}}$, 
	total energy change $\Delta E$ with respect to a polaron position in the domain interior, 
	relative polaron density $n(s)/n_0$ as a function of temperature ($n_0$: density in the domain interior), 
	in-plane averaged and smoothened electronic potential $\langle V\rangle$, 
	local direction-averaged strain, and octahedral tilt profiles around the three cartesian axes 
	for small electron polarons near these domain walls are 
	shown in \textbf{Figure}~\ref{fig:energy_dens_pol}.
	The density was calculated with Eq.~(\ref{eq:conc}), the ionic polarization with Eq.~(\ref{eq:pol}).
}
%\begin{figure}[htb]
%	\includegraphics[width=0.49\textwidth]{energies_SEP_71DW.pdf}\\
%	\includegraphics[width=0.49\textwidth]{density_SEP_71DW.pdf}
%	\caption{\label{fig:energies_SEP_71DW}\new{Energy (top) and density (bottom) profile for small electron polarons at the 71\textdegree DW}.}
%\end{figure}
\new{The local strain with polaron in the lowest-energy site appears to couple to existing strain variations at the walls, 
possibly also to tilt variations. The local electronic potential in the pristine system, before polaron formation, 
 exhibits a minimum at or near the lowest-energy polaron site.
	}
\begin{figure}[htb]
	\includegraphics[width=0.33\textwidth]{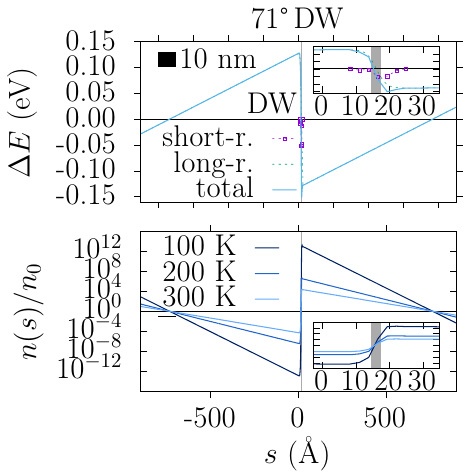}
	\includegraphics[width=0.33\textwidth]{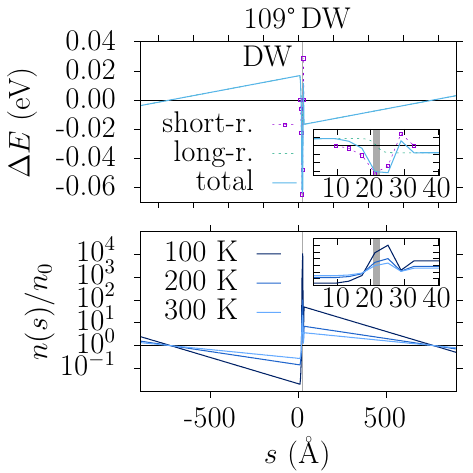}
	\includegraphics[width=0.33\textwidth]{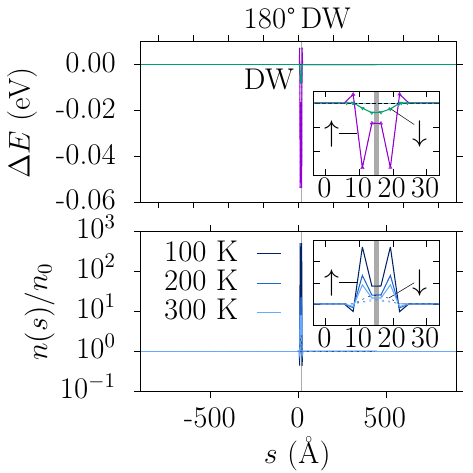}
	\caption{\label{fig:longrange_pot_dens}\new{potential energy profile and polaron density at the 71\textdegree\ (left), 109\textdegree\ (center), and the 180\textdegree\ DW (right),
	assuming short-circuit conditions. The insets show a zoom into the domain wall.}
	}
\end{figure}
\new{The trapping energies for small electron polarons at the domain walls are listed in \textbf{Table}~\ref{tab:trapping_energies}.}
\begin{table}
	\caption{\label{tab:trapping_energies} Computed corrected trapping energies for electron polarons at ferroelectric domain walls in eV.
	}
	\begin{tabular}{l | c | c }
		&  \multicolumn{2}{c}{$E_{\mathrm{trap}}$ (eV)}  \\	
		DW   & 160-atom supercell & 240-atom supercell \\			\hline
		71\textdegree &    -0.063  &  -0.051  \\
		109\textdegree &    -0.075  & -0.065  \\
		180\textdegree &   -0.044  &  -0.054  \\		
	\end{tabular}
\end{table}
%%%%%%%%%%%%%%%%%%%%%%%%%%%%%%%%%%%%%%%%%%%%%%%%
%\subsubsection{\new{109\textdegree\ and 180\textdegree\ domain wall}}
%%%%%%%%%%%%%%%%%%%%%%%%%%%%%%%%%%%%%%%%%%%%%%%
%\new{The energy profiles, relative densities, and ionic polarization profiles for small electron polarons near the 109\textdegree\ domain wall and the 180\textdegree\ domain wall are shown in \textbf{Figure}~\ref{fig:energy_dens_pol}.
\new{
We observe two different trap depths for the two 180\textdegree\ domain walls even though the two domain walls are equivalent by symmetry; 
this is because the symmetry operation relating first and second wall exchanges the magnetic sublattices. Here only polarons on spin-down sites were considered, 
the profile for polarons of the other spin was then obtained based on symmetry.  
%The energy profile for the spin-up sites is shifted by one domain-wall spacing.
The computed trapping energies (-51\,meV for the 71\textdegree, -65\,meV for the 109\textdegree, and -54\,meV for the 180\textdegree\ domain wall) are 
similar to each other.
The 180\textdegree\ domain wall differs from the 71\textdegree\ and 109\textdegree\ walls in that it attracts small polarons in a plane one layer apart from the layer closest to the wall, 
and, interestingly, each of the two equivalent 180\textdegree\ walls selectively traps essentially only polarons with a specific spin. 
The trapping energies were 
%extrapolated to infinite $k$-point density and 
linearly corrected an for underestimated ferroelectric energy gain with PBEsol as the reference, 
see Table~\ref{tab:compare_xc_functionals} and section \ref{sec:methods:e_at_DW}.
The short-range potential variations at the doman walls are superimposed on a large-scale potential landscape that depends on the boundary conditions, 
for example, open- or closed-circuit conditions, surface termination, etc.
To estimate the density of excess electrons at the walls, one should consider both short-range and long-range potential. 
In a previous work \cite{koerbel:2020:photovoltage}, the long-range potential at 71\textdegree\ and  109\textdegree\ domain walls 
was found to consist in a potential step at the wall and a potential slope in the domains that depends on the boundary conditions. 
For zero potential slope (zero electric field in the domains), only the short-range potential acts on the polarons, 
and their density should be determined by Eq.~\ref{eq:conc}, as depicted in Figure~\ref{fig:energy_dens_pol}.
For closed-circuit conditions, the potential step at the domain walls should create an electric field in the domains. 
This case is considered in \textbf{Figure}~\ref{fig:longrange_pot_dens}, assuming a domain-wall spacing of 150\,nm \cite{seidel:2010:domain,seidel:2011:efficient} 
and a polaron density in the dilute limit ($n$ and $n_0$ very small).
}
%%%%%%%%%%%%%%%%%%%%%%%%%%%%%%%%%%%%%%%%%%%%%%%%%%%%%%%%%%%%%%%%%%%
\subsubsection{\new{Polaron hopping at domain walls}}
%%%%%%%%%%%%%%%%%%%%%%%%%%%%%%%%%%%%%%%%%%%%%%%%%%%%%%%%%%%%%%%%%%%
%
\new{For the 71\textdegree\ domain wall, polaron hopping was investigated along and across the domain wall.}
When starting from a delocalized-electron configuration \new{at the 71\textdegree\ domain wall}, 
a small electron polaron forms spontaneously in the plane labeled ``1'' 
(on the right-hand side of the wall, if the net polarization points to the right, see Figure~\ref{fig:hopping_paths}(b)),
which is a metastable site.
The most stable site is inside the domain wall, in the plane labeled ``0''.
\begin{figure*}[h]
 %\includegraphics[width=0.45\textwidth]{DW_2NN_hops.pdf}
 %\includegraphics[width=0.45\textwidth]{DW_2NN_low_E_hops.pdf}
 %\includegraphics[width=0.6\textwidth]{DW_NN_2NN_E_vs_coord.pdf}
%	\hspace{0.09\textwidth}\includegraphics[width=0.9\textwidth,height=0.3\textwidth,keepaspectratio=FALSE]{CONTCAR_new_labels_and_hops_2.pdf}
 \includegraphics[width=0.7\textwidth]{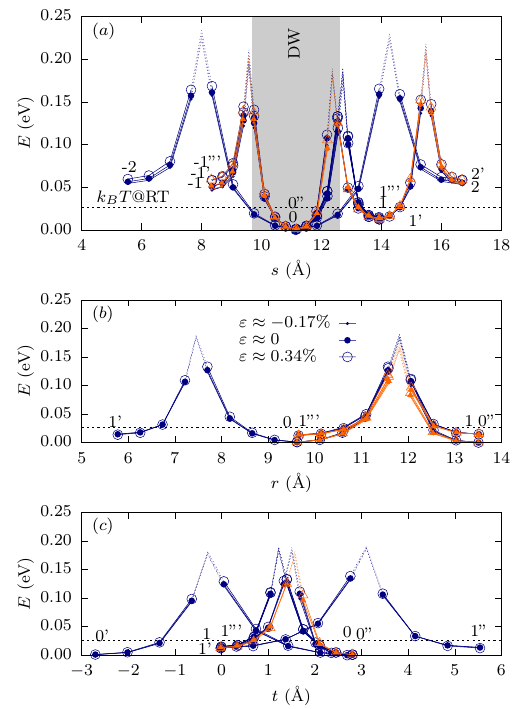}
 \includegraphics[width=0.25\textwidth]{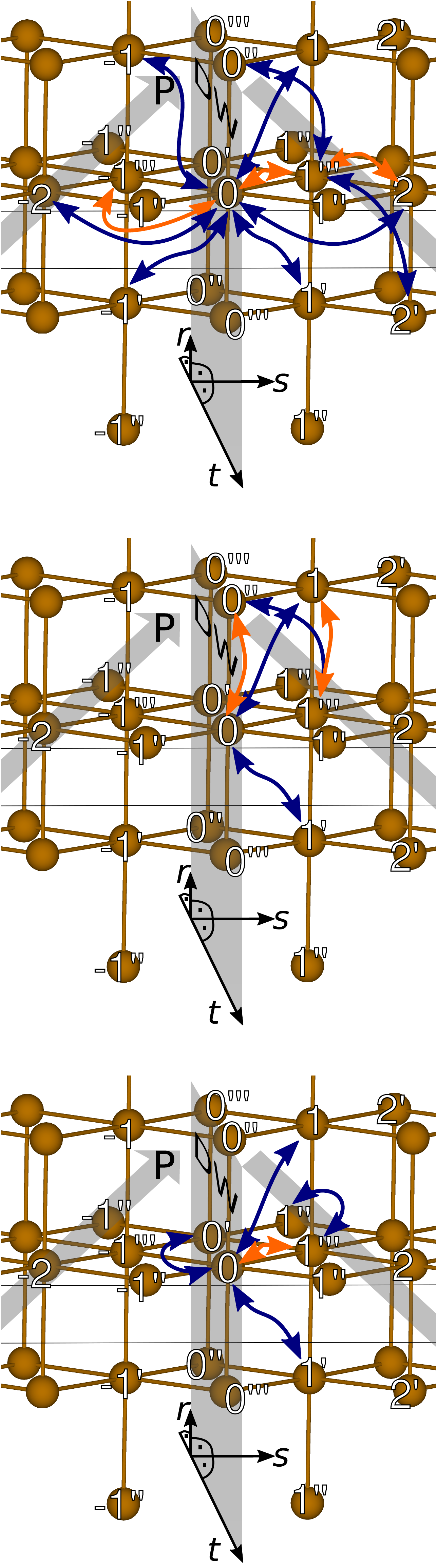}
%	\caption{\label{fig:DW_hops}Top: second-nearest neighbor hops from Fe18, the metastable site. Center: hops from Fe5, the stable site. 
%	Bottom: both.}
	\caption{\label{fig:DW_hops}Energy barriers of polaron hops at the 71\textdegree~domain wall. 
	(a) All hops with a component across the wall, (b) and (c): selected hops with components along the wall. 
	Orange triangles: nearest-neighbor, dark-blue circles: second-nearest-neighbor hops. 
	Numbers indicate the Fe positions in Figure~\ref{fig:hopping_paths}(b). 
	Hop directions were projected on (a) the $s$ coordinate, (b) the  $r$ coordinate, and (c) the $t$ coordinate, see Figure~\ref{fig:hopping_paths}(b).}
\end{figure*}
%%%%%%%%%%%%%%%%%%%%%%%%%%%%%%%%%%%%%%%%%%%%%%%%%%%%%%%%%%%%%%%%%%%
%\subsubsection{Nearest neighbor hops, spin-flip needed}
%%%%%%%%%%%%%%%%%%%%%%%%%%%%%%%%%%%%%%%%%%%%%%%%%%%%%%%%%%%%%%%%%%%
%\begin{figure}[h]
% \includegraphics[width=0.45\textwidth]{DW_NN_hops.pdf}
% \includegraphics[width=0.45\textwidth]{DW_NN_low_E_hops.pdf}
%	\caption{\label{fig:DW_hops}Top: Nearest neighbor hops from Fe18, the metastable site. Bottom: hops from Fe5, the stable site.}
%\end{figure}
%
\textbf{Figure}~\ref{fig:DW_hops} shows the energy barriers for electron hops at the 71\textdegree~domain wall. 
Only \new{some representative low } barriers in each direction are shown. 
\new{The computed 2NN energy barriers are listed in \textbf{Table}~\ref{tab:barriers_DW}}. 
All barriers are depicted and tabulated in the Supporting Information.
The hops are projected on the symmetry-adapted coordinates %(see Figure~\ref{fig:DW_hops}) 
$s$ [$\perp$ wall, (a)], $r$ [within the wall, (b)], and $t$ [within the wall, $\perp$ $\vec{P}$, (c)]. 
For hops between the two trap states in the planes ``0'' and ``1'', the barrier is \new{about 0.2~eV; 
for hops out of the wall  (from plane ``0'' or ``1'' to plane ``-1'',``2'', or ``-2'') the barrier is about 10 to 20\% larger.}
%For hops within the domain wall [(b) and (c)] there are pathes available with barriers similar to the thermal energy at room temperature. 
The barriers are invariant with respect to translation by a lattice vector parallel to the wall. 
Like in the bulk, the barriers \new{slightly increase under tensile strain}.
The lowest barriers for NN and 2NN hops are similar in size.
% 2NN hops type r:
% -0.17      30.65986666666667  # (35.1826+28.8213+27.9757)/3.0
% 0          31.676666666666666 # (37.3722+27.8498+29.808)/3.0
% 0.34       33.291533333333334 # (37.3883+31.236+31.2503)/3.0
% extrapolated: 34.7442457340091

% 2NN hops type t2:
% -0.17           16.1782
% 0               25.4104
% 0.34            27.2722
% extrapolated:  28.94705031638

% 2NN escape hops:
%-0.17     56.996 # (57.8417+ 56.1503 )/2.0
%0         61.7786 # (65.2156+58.3416)/2.0
%0.34      64.75555 # (68.3597+61.1514)/2.0
% extr.: 67.4335743040861

\begin{table}
	\caption{\label{tab:barriers_DW} Computed 2NN energy barriers $E_a$\new{, electron mobility $\mu_e^{\mathrm{DW}}$, and diffusion constant $D^{\mathrm{DW}}$} for small electron hopping in different directions at 71\textdegree~domain walls (DW) in BiFeO$_3$ 
	for different strains and extrapolated to the experimental cell volume at room temperature. 
	The estimated uncertainty is 10\% or $\approx$20\,meV.
	The extrapolated barriers were used to calculate the mobility and the diffusion constant with Eq.~(\ref{eq:mu_DW}).}
	\begin{tabular}{l | c | c | c | c | c | c | c  }
		\hline
			       & \multicolumn{3}{c|}{$E_a$ (eV)}         & \multicolumn{3}{c|}{ }  &         \\
		               &   double 2NN,   & 2NN,       &   2NN, escape           & \multicolumn{3}{c|}{$\mu_e^{\mathrm{DW}}$ (cm$^2$/Vs)}   &   \\
		$\varepsilon$  &    $r$ or $t2$  &   $t1$     & from DW  & $\mu_{e,r}^{\mathrm{DW}}$   & $\mu_{e,t}^{\mathrm{DW}}$ & $\mu_e^{\mathrm{DW}}$   &  $D^{\mathrm{DW}}$  (cm$^2$/s) \\\hline
		$-0.17$\%      &    0.19        &  0.19       &  0.22    & -- & --& &  \\
		 0             &    0.20        &  0.19       &  0.23    & -- & --& &  \\
		 0.34\%        &    0.20        &  0.20       &   0.23   & -- & --& &  \\
		 %0.65\% (extrapol.) &   0.20     & 0.20       &   0.24   & & & $0.2\dots 1 \cdot 10^{-3}$  &   $0.6 \dots 3 \cdot 10^{-5} $            \\\hline
             1.9\% (extrapol.) &    0.21        &  0.21       &   0.25   & $0.4\dots 2 \cdot 10^{-4}$  & $0.3\dots 1 \cdot 10^{-4}$ & $0.7\dots 4 \cdot 10^{-4}$  &   $0.2 \dots 1 \cdot 10^{-5} $            \\\hline
	\end{tabular}
\end{table}

\begin{figure}[htb]
	\includegraphics[width=0.6\textwidth]{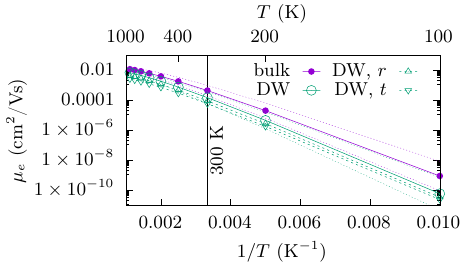}
	\caption{\label{fig:mu_vs_T}\new{Computed small electron polaron mobility in bulk BiFeO$_3$ and inside ferroelectric 71\textdegree\ domain walls (DW) 
	as a function of temperature. The lines serve as a guide to the eye. The dashed lines delineate the error margin.}}
\end{figure}

\new{
The computed two-dimensional electron mobility inside the domain-wall plane is $\mu_e^{\mathrm{DW}}\approx 2\cdot 10^{-4}$\,cm$^2$/Vs, 
the computed two-dimensional diffusion constant inside the domain wall is $D^{\mathrm{DW}}\approx 4\cdot 10^{-6}$\,cm$^2$/s. 
%$\mu_e$ and $D$ have an uncertainty of about a factor of two.
Assuming 10\% uncertainty in the barriers, one obtains $\mu_e^{\mathrm{DW}}\approx$ 0.7 to 4 $\cdot 10^{-4}$\,cm$^2$/Vs and $D^{\mathrm{DW}}\approx$ 0.2 to 1 $\cdot 10^{-5}$\,cm$^2$/s.}
%
%%%%%%%%%%%%%%%%%%%%%%%%%%%%%%%%%%%%%%%%%%%%%%%%
%\subsubsection{\new{180\textdegree\ domain wall}}
%%%%%%%%%%%%%%%%%%%%%%%%%%%%%%%%%%%%%%%%%%%%%%%
%\new{The energy profile for small electron polarons near the 180\textdegree\ domain wall is shown in \textbf{Figure}~\ref{fig:energies_SEP_180DW}.
%}
%\begin{figure}[htb]
%	\includegraphics[width=0.49\textwidth]{energies_SEP_180DW.pdf}
%	\caption{\label{fig:energies_SEP_180DW}\new{Energy profile for small electron poalrons at the 180\textdegree DW}.}
%\end{figure}
\new{The temperature dependence of the computed mobility is depicted in \textbf{Figure}~\ref{fig:mu_vs_T}. 
The mobility is dominated by the exponential decay originating in the hopping barrier.
	}
%%%%%%%%%%%%%%%%%%%%%%%%%%%%%%%%%%%%%%%%%%%%%%%%%%%%%%%%
\section{Discussion}
%%%%%%%%%%%%%%%%%%%%%%%%%%%%%%%%%%%%%%%%%%%%%%%%%%%%%%%%
\new{All three ferroelectric domain walls, the 71\textdegree, the 109\textdegree, and the 180\textdegree\ walls, 
trap small polarons on a short range, with a trap depth of about 50 to 65 meV. 
Depending on the boundary conditions, such as applied voltage or surface termination, 
a long-range potential exists in the case of the 71\textdegree\ and 109\textdegree\ domain wall,
which originates from a potential step at the domain walls and can create an electric field in the domains, 
whose strength depends on the boundary conditions: the field should be maximal 
for closed-circuit conditions and zero for open-circuit conditions.  
For symmetric 180\textdegree\ domain wall, there should be no electric field. 
The electric field, if present, dominates the potential energy landscape for small polarons
if the potential step at the wall is large compared to the trapping energy,
which is the case for the 71\textdegree\ domain wall (potential  step: 0.13~eV, trap depth: 51\ meV).
Assuming a resolution of 10 nm for spatially resolved conductivity measurements at domain walls, 
the increase in polaron density at the domain wall due to the short-range trap should be hidden behind the large density increase due to the long-range 
trap originating in the potential step at the wall. 
In the case of open-circuit conditions, the polaron density and conductivity increase at the wall should originate from the short-ranged 
trapping profile. In this case, only a moderate increase in polaron density and conductivity of about one order of magnitude is to be expected 
if measured on a ten-nanometer scale, and only at low temperatures, e.g. at 100\,K. 
The electrical conductivity in the domain interior and at the domain walls will be further influenced by atomic point defects, 
which are not considered here.}

Hall measurements for hematite resulted in electron mobilities of about 0.01 to $>$1~cm$^2$/Vs \cite{morin:1951:electrical,vandaal:1967:hall,zhao:2011:electrical},
\new{which is one to two orders of magnitude higher than the here computed mobility in bulk BiFeO$_3$ of $10^{-4}$ to $10^{-3}$~cm$^2$/Vs. 
Considering that the Fe-Fe distance in BiFeO$_3$ is larger than in hematite, and that NN hops are spin-forbidden in BiFeO$_3$, 
a lower mobility is to be expected in BiFeO$_3$.
In order to obtain the same mobility for BiFeO$_3$ as in hematite, the energy barriers would need to be about 40\% smaller,
near  0.1~eV instead of 0.2~eV.
At elevated temperatures,   
%It does not follow that the computed energy landscape is unrealistic in itself. 
 a combination of electron tunneling through the top of the barrier, 
smearing out of the traps due to thermal motion, incomplete self-trapping, and occupation of excited self-trapped states should effectively reduce the barriers. 
At low temperature, the effective barriers should approach those computed here.
}
\new{The here calculated hopping barriers ($\approx 0.2$\,eV) are similar to those obtained for PbTiO$_3$ \cite{ghorbani:2022:self}, 
so that the activation energy for the intrinsic electron mobility in BiFeO$_3$ should be comparable to that in PbTiO$_3$. 
}

Experimentally determined activiation energies of electron conduction in BiFeO$_3$ are typically \new{in a similar range or larger}  
(between about 0.1 and 1.3~eV \cite{ramirez:2017:comprehensive,das:2018:impedance,kolte:2015:impedance,perejon:2013:electrical,pintilie:2009:orientation}) 
than the here computed barrier of about \new{0.2\,eV}.
\new{The lower measured barriers could hence originate from electron polaron hopping, 
while it seems likely that the experimentally determined higher barriers of several hundred meV originate from defects other than pristine neutral bulk or ferroelectric domain  
walls, for example from point defects, such as oxygen or bismuth vacancies, or from interface effects, such as Schottky barriers.}

In Refs.~\cite{farokhipoor:2011:conduction,yao:2012:pyroelectric} an activation barrier of $\approx 0.7$~eV was found for $n$-type conduction, 
which is about the energy difference between the electron polaron level and the conduction band minimum computed here.
In principle, this activation barrier of 0.7~eV could originate from emission of electrons from the polaron state to the conduction band.
This would mean that transport by emission from polarons into delocalized conduction states and subsequent ballistic transport would be more efficient 
than polaron hopping, inspite of the much higher energy barrier that must initially be overcome.
However, there are reasons to believe that the lifetime of thermalized electrons in the conduction band at room temperature should be small: 
\new{The strong Fe-$d$ contribution} to the states at the conduction band minimum and the small size of the electron polaron indicate that electron-phonon coupling should be large, 
which should lead to strong electron-phonon scattering. 
First-principles molecular dynamics simulations yield a polaron formation time of less than a picosecond,
see \new{Supporting Information, similar to hematite \cite{cheng:2022:photoinduced}. 
For hematite, transient extreme-ultraviolet spectroscopy yielded ``signatures characteristic of electron localization by small polarons [...] within 100\,fs following photoexcitation'' \cite{carneiro:2017:excitation}.}
%and Ref.~\cite{koerbel:2024:erratum}.}
\new{Hence, it seems more likely that the measured barrier of 0.7\,eV belongs to a defect or interface.}

Once a small electron polaron has formed, it is unlikely to transform into a delocalized conduction-band electron at room temperature and below since the energy distance between the polaron level and the conduction-band minimum 
 of \new{$|\varepsilon_p-E_{\mathrm{CBM}}|\approx 0.7$\,eV} is much larger than the polaron hopping barrier of \new{$E_a\approx 0.2$\,eV}.
\new{It seems most likely that %at least for thermalized electrons at temperatures $\lesssim$ room temperature, 
small electron hopping should be the dominant intrinsic $n$-type transport process at room temperature and below.
%, and that in the reported experimental measurements of the electrical conductivity, 
%the barrier of $\approx 0.2$\,eV for electron polaron hopping is 
%hidden behind larger barriers associated with emission from defects or interfaces other than pristine neutral ferroelectric domain walls.
}

%Alternatively, it is possible that measured electron transport is typically dominated by different defect states, which hide the intrinsic mobility,
%and that the 0.7~eV activation energy belong to one of the defect states.

%\new{The approach taken here allows to determine the intrinsic mechanism of $n$-type conduction, once 
\new{Here it is assumed that excess electrons have been brought into BiFeO$_3$ in some way. 
For example, charge carriers may originate from point defects, such as oxygen vacancies or aliovalent dopants or impurities, 
from photoexcitation resulting in bound or dissociated electron-hole pairs, 
or from an electrical contact. 
%The approach taken here can predict the intrinsic charge carrier mobility, 
%but it cannot predict the \new{absolute} charge-carrier density,
%since the latter depends on the sample fabrication and history and the experimental conditions, 
%and it can hence not predict the absolute electrical conductivity, 
%but it can predict the temperature dependence of the electrical conductivity and the effect of deviations from the perfect 
%crystal structure. 
%External processes of emitting and trapping electrons by defects are not part of the intrinsic electrical conduction mechanism, 
%and considering the effects of all possible defects is here deferred to future work. 
%However, a particular defect (or desired structure modification) is considered here, namely 
%the neutral ferroelectric 71\textdegree~domain wall.
}

%%%%%%%%%%%%%%%%%%%%%%%
\section{Summary and conclusion}
%%%%%%%%%%%%%%%%%%%%%%%
\new{Summarizing, the domain wall planes act as traps for small electron polarons; %
%, with a trap depth of about two to three times $k_B T$ at room temperature ($\approx$\,$70\pm10$\,meV).
trapping energies of -51\,meV (71\textdegree\ wall), -75\,meV (109\textdegree\ wall), and -54\,meV (180\textdegree\ wall) were obtained here. 
The origin of the trapping is likely a coupling of the polaron deformation to existing strain and possibly tilt variations at the domain walls.
These trap depths are about two times $k_B T$ at room temperature and hence near the boundary between deep and shallow levels.
Assuming $3k_BT$ as the boundary between shallow and deep levels \cite{wikipedia:2025:shallow},
the domain walls act as shallow traps at and above room temperature and become deep traps 
for temperatures below $\approx 200$ to $250$ K.
The atomic and magnetic structure of the 180\textdegree\ domain walls couple in such a way that the walls selectively trap polarons depending on their spin.
Pristine domain walls, especially 71\textdegree\ domain walls, can accumulate polarons in a region of tens of nanometers 
up to densities that are several orders of magnitude larger than in the bulk, depending on the boundary conditions. 
This is due to a potential step in the wall that can create a long-ranged electric field. 
For the 109\textdegree\ domain wall, a maximum polaron density of less than one order of magnitude more than in the bulk is possible 
based on the computed potential step, at symmetric 180\textdegree\ walls there should be no such long-range electric field. 
The experimentally measured increase in electrical conductivity at the domain walls should essentially originate in a local increase in polaron density at the walls.}

\new{For the first time, energy barriers for small electron polaron hopping in BiFeO$_3$ were determined using density-functional theory. 
Also for the first time, electron polaron hopping at a ferroelectric domain wall was modeled from first principles.} 
Hopping in perfect, pristine BiFeO$_3$ and at pristine, neutral ferroelectric 71\textdegree~domain walls was considered.

%Both nearest-neighbor and second-nearest neighbor hops were insvestigated, since due to the magnetic structure of BiFeO$_3$, 
%the nearest-neighbor hops are spin-forbidden.
In both bulk and domain-wall plane, the calculated energy barriers \new{for hopping are about 0.2\,eV
and slightly increase under tensile strain}.
%The nearest-neighbor and second-nearest neighbor hopping barriers are similar in size,
%so that transport should be dominated by second-nearest neighbor hops, which are spin-allowed.
%The barriers are not strongly direction dependent in bulk and within the domain-wall plane.
\new{The calculated hopping barrier for electrons in BiFeO$_3$ is similar, and the resulting activation barrier for the electron mobility should be similar
 to, e.g., PbTiO$_3$.} %, suggesting that BiFeO$_3$ could be a promising material for optoelectronic applications that benefit from high charge carrier mobility, such as water splitting.}
Using non-adiabatic Marcus theory, an electron mobility of 
$\mu_e \approx 10^{-4}$ to $10^{-3}$\,cm$^2$/Vs %and $\mu_e^{\mathrm{DW}}\approx 0.8$\,cm$^2$/Vs for bulk and within the domain-wall plane, 
and  a computed diffusion constant of $D\approx$0.2 to 3$\cdot 10^{-5}$\,cm$^2$/s 
are obtained at room temperature both in the bulk and in the domain-wall planes. 
%The domain-wall contribution to $n$-type electronic conductivity should arise 
%from hopping of small electron polarons trapped by the domain walls. 
%Their transport should be confined to the domain-wall plane at temperatures below about twice room temperature,
%and it should be thermally activated near room temperature.
%Escape of electrons from the wall should require two to three times higher temperatures.
%
\\ \ \\
\new{Concluding,} % these findings indicate that
(a) experimentally found activation barriers of several hundred meV for electrical conductivity in BiFeO$_3$
should have an origin other than electron polaron hopping in pristine BiFeO$_3$ or electron polaron escape from pristine neutral ferroelectric 71\textdegree~domain walls. 
\new{To identify the origin of these large energy barriers is here deferred to future work.
This will be a considerable task, but it should be undertaken to fully understand the limits of electronic conductivity in BiFeO$_3$ and 
other transition-metal oxides.}
\\ 
(b) The effect of pristine, neutral ferroelectric domain walls on the $n$-type conductivity at room temperature consists in accumulating excess electrons 
and confining their motion to the domain-wall plane. 
In the Drude picture (conductivity$\sim$carrier density$\cdot$mobility), the electron density \new{at} the wall is enhanced compared to the bulk, 
%the mobility along the wall (hopping barrier) is about the same as in bulk, 
the mobility perpendicular to the wall is \new{slightly} reduced compared to the bulk.
%The total $n$-type conductivity should be composed of both bulk and domain-wall contributions, 
%where the bulk contribution should be approximately isotropic.
This is in line with experimental observations that conductivity is enhanced at ferroelectric domain walls  
and in directions parallel to the domain walls.
\new{At temperatures up to room temperature, pristine ferroelectric 71\textdegree\ domain walls can in principle 
exhibit a carrier density and an electrical conductivity up to several orders of magnitude larger than in the bulk, 
depending on the boundary conditions, due to a potential step of about 0.13~eV in the wall. 
For the 109\textdegree\ domain wall, assuming a spatial resolution of ten nanometers, 
a carrier density and conductivity enhancement by up to one order of magnitude is possible, 
due to a potential step of about 17\,meV.
For the symmetric 180\textdegree\ domain wall, the carrier density is enhanced by far less than one order of magnitude, 
when averaged on a ten-nm scale. 
Accumulation of point defects other than electron polarons, not considered here, 
should further modify the potential and carrier density profile of the ferroelectric domain walls.}
%for example....check also temperature of experimental measurements}

%\clearpage

\medskip
\textbf{Supporting Information} \par %Please delete the Suppporting Information statement if it is not applicable. Please supply Supporting Information in another file. Supporting information should not be provided in .tex format
Supporting Information is available from the Wiley Online Library or from the author.

% Acknowledgements
\medskip
\textbf{Acknowledgements} \par %delete if not applicable))
%%%%%%%%%%%%%%%%%%%%%%%%%%%%%%%%%
S.~K. thanks Michele Reticcioli for recommending the OMC method. %and
% Johannes Steinmetzer for helpful discussions about transition state theory.}
This project  has received  funding  
\new{from  the European Union's Horizon 2020 research and innovation programme under the Marie Sk\l{}odowska-Curie Grant Agreement No. 746964, 
%from a postdoc stipend of the University of Jena, Germany, 
from a scholarship programme at the University of Jena within the Thuringian Scheme for Young Female Researchers and Young
Female Artists funded by the Free State of Thuringia,}
and from the Volkswagen Stiftung (Momentum) through the project ``dandelion''. 
Computational resources and support were supplied by 
the HPC cluster ARA of the University of Jena, Germany.
Figures were made using {\sc{Vesta}} \cite{momma:2011:vesta} and gnuplot.

%%%%%%%%%%%%%%%%%%%%%%%%%%%%%%%%%%%%%

% References
\medskip

% Use the following code if you wish to generate your bibliography with BibTeX;
% replace the string "MSP-template" below with the name(s) of
% the BibTeX data base(s) you want to use.
% The resulting bibliography-output (the content of the .bbl file)
% must be pasted back into this file before submission.
% Please also include your BibTeX data base file(s) in your submission
% so that we can re-run BibTeX if necessary.
%
\bibliographystyle{MSP}
\bibliography{jabbr,all}
%\bibliography{/media/koerbel/Elements/Literature/LatexBib/jabbr,/media/koerbel/Elements/Literature/LatexBib/all}

%\textbf{References}\\

%%%%%%%%%%%%%%%%%%
%\newpage
%\section*{Supp. Inf.}
%%%%%%%%%%%%%%%%%%
%
% Table of contents entry should be 50 - 60 words long
% Image should be 55 mm broad and 50 mm high or 110 mm broad and 20 mm high

%\begin{figure}
%\textbf{Table of Contents}\\
%\medskip
%  %\includegraphics{toc-image.png}
%  \includegraphics[width=0.25\textwidth]{2NN_path_PARCHG_00.pdf}
%  \medskip
%  \caption*{ToC Entry}
%\end{figure}

\end{document}

\pagestyle{fancy}
%\rhead{\includegraphics[width=2.5cm]{vch-logo.png}}
\rhead{ }

%\title{Energy barriers for small electron polaron hopping in bismuth ferrite from first principles\\
\title{\new{Energy profile and hopping barriers for small electron polarons at ferroelectric domain walls} in bismuth ferrite from first principles\\
Supporting Information}

\maketitle

% Author: Please give full first and last names for authors and include * after the name of all corresponding authors
\author{Sabine K\"{o}rbel*}
% Dedication

\dedication{}

% Affiliations: Please provide adacemic titles (Prof. or Dr.) for all authors where applicable, and include an institutional email address for all corresponding authors
\begin{affiliations}
\jena \\
Email Address: skoerbel@uni-muenster.de
\end{affiliations}

% Keywords: Please provide a minimum of three and a maximum of seven keywords, separated by commas

\keywords{polarons, ferroelectrics, domain walls, density-functional theory, electronic transport}

%%%%%%%%%%%%%%%%%%%%%%%%%%%%%%%%%%%%%%%%%%%%%%%%%%%%%%%%%%%%%%%%%%%
\section{Energy barrier for electron polaron hopping in bulk BiFeO$_3$}
%%%%%%%%%%%%%%%%%%%%%%%%%%%%%%%%%%%%%%%%%%%%%%%%%%%%%%%%%%%%%%%%%%%
\begin{figure*}[htb]
 \includegraphics[width=0.2\textwidth]{PARCHG_img4.pdf}
 \includegraphics[width=0.59\textwidth]{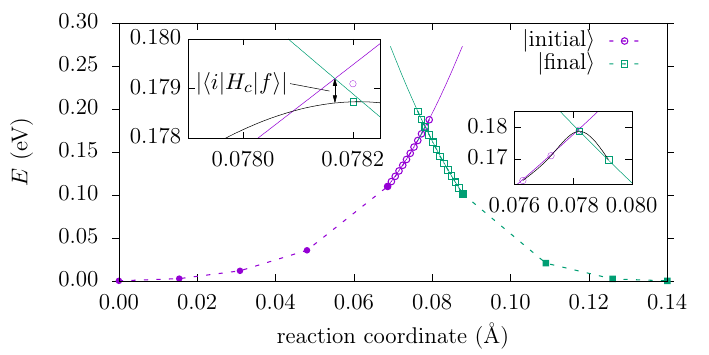}
 \includegraphics[width=0.2\textwidth]{PARCHG_img5.pdf}
	\put(-500,80){$|\mathrm{initial}\rangle$}
	\put(-50,80){$|\mathrm{final}\rangle$}
\caption{\label{fig:energy_NEB_bulk_hires}Energy profile along the reaction coordinate (the maximum atomic displacement) for a strain-free 2NN hop in bulk with $\alpha\approx 90\degree$. 
Filled symbols: NEB, empty symbols: linear interpolation.
The black solid line is a cubic spline.}
\end{figure*}
\new{From a cubic fit of the nearly-diabatic branches, one obtains a coupling matrix element of $\langle\psi_i|H_c|\psi_f\rangle\approx -0.1$ to $-1$\,meV.}
%%%%%%%%%%%%%%%%%%%%%%%%%%%%%%%%%%%%%%%%%%%%%%%%%%%%%%%%%%%%%%%%%%%
\section{Energy barriers for all hops at the 71\textdegree~domain wall}
%%%%%%%%%%%%%%%%%%%%%%%%%%%%%%%%%%%%%%%%%%%%%%%%%%%%%%%%%%%%%%%%%%%

\begin{table}[htb]
	\caption{\label{tab:all_barriers_DW_2NN} All computed 2NN energy barriers $E_a$ in meV (maximal, for hop to, for hop fro) for small electron polaron hopping at 71\textdegree~domain walls in BiFeO$_3$ 
	for different strains.}
	\begin{tabular}{l| l| l | l | l | l}
		strain $\varepsilon$ &  2NN  hop                       &  max. barrier  &  barrier $\to$ & barrier $\leftarrow$ & along \\\hline
-0.17\% &$ 1'''         \leftrightarrow          1'' $ &   187 	 & 187 	 & 187   &  $t $                \\
-0.17\% &$ 1'''         \leftrightarrow          0'' $ &   193 	 & 180 	 & 193   &  $s,r,t $   \\
-0.17\% &$ 1'''         \leftrightarrow          2'  $ &   191 	 & 191 	 & 151   &  $s,r,t $   \\
-0.17\% &$ 0            \leftrightarrow          1   $ &   190 	 & 190 	 & 176   &  $s,r,t $   \\
-0.17\% &$ 0            \leftrightarrow          -1  $ &   195 	 & 195 	 & 149   &  $s,r,t $   \\
-0.17\% &$ 0            \leftrightarrow          -1' $ &   188 	 & 188 	 & 143   &  $s,r,t $   \\
-0.17\% &$ 0            \leftrightarrow          -2  $ &   224 	 & 224 	 & 171   &  $ s$       \\
-0.17\% &$ 0            \leftrightarrow          1'  $ &   191 	 & 191 	 & 177   &  $s,r,t $   \\
-0.17\% &$ 0            \leftrightarrow          0'  $ &   188 	 & 188 	 & 188   &  $t $       \\
-0.17\% &$ 0            \leftrightarrow          2   $ &   222 	 & 222 	 & 169   &  $s $       \\
 0\%    &$ 1'''         \leftrightarrow          1'' $ &   192 	 & 192 	 & 192   &  $t $              \\
 0\%    &$ 1'''         \leftrightarrow          0'' $ &   196 	 & 182 	 & 196   &  $s,r,t $   \\
 0\%    &$ 1'''         \leftrightarrow          2'  $ &   201 	 & 201 	 & 159   &  $s,r,t $   \\
 0\%    &$ 0            \leftrightarrow          1   $ &   197 	 & 197 	 & 183   &  $s,r,t $   \\
 0\%    &$ 0            \leftrightarrow          -1  $ &   204 	 & 204 	 & 154   &  $s,r,t $   \\
 0\%    &$ 0            \leftrightarrow          -1' $ &   199 	 & 199 	 & 149   &  $s,r,t $   \\
 0\%    &$ 0            \leftrightarrow          -2  $ &   232 	 & 232 	 & 176   &  $ s$       \\
 0\%    &$ 0            \leftrightarrow          1'  $ &   197 	 & 197 	 & 184   &  $s,r,t $   \\
 0\%    &$ 0            \leftrightarrow          0'  $ &   195 	 & 195 	 & 194   &  $t $       \\
 0\%    &$ 0            \leftrightarrow          2   $ &   229 	 & 229 	 & 175   &  $s $       \\
 0.34\% &$ 1'''         \leftrightarrow          1'' $ &   195 	 & 195 	 & 195   &  $t $       \\
 0.34\% &$ 1'''         \leftrightarrow          0'' $ &   199 	 & 186 	 & 199   & $s,r,t $   \\
 0.34\% &$ 1'''         \leftrightarrow          2'  $ &   208 	 & 208 	 & 162   & $s,r,t $   \\
 0.34\% &$ 0            \leftrightarrow          1   $ &   200 	 & 200 	 & 186   & $s,r,t $   \\
 0.34\% &$ 0            \leftrightarrow          -1  $ &   212 	 & 212 	 & 155   & $s,r,t $   \\
 0.34\% &$ 0            \leftrightarrow          -1' $ &   205 	 & 205 	 & 148   & $s,r,t $   \\
 0.34\% &$ 0            \leftrightarrow          -2  $ &   236 	 & 236 	 & 177   & $ s$       \\
 0.34\% &$ 0            \leftrightarrow          1'  $ &   201 	 & 201 	 & 188   & $s,r,t $   \\
 0.34\% &$ 0            \leftrightarrow          0'  $ &   197 	 & 197 	 & 197   & $t $       \\
 0.34\% &$ 0            \leftrightarrow          2   $ &   233 	 & 233 	 & 175    & $s $   
	\end{tabular}
\end{table}

\begin{table}[htb]
	\caption{\label{tab:all_barriers_DW_NN} All computed NN energy barriers $E_a$ in meV (maximal, for hop to, for hop fro) for small electron polaron hopping at 71\textdegree~domain walls in BiFeO$_3$ 
	for different strains.}
	\begin{tabular}{l| l| l | l | l}
strain $\varepsilon$ &  NN  hop          &  max. barrier  &  barrier $\to$ & barrier $\leftarrow$ \\\hline
-0.17\% & $ 1''' \leftrightarrow         1      $ &  148 	 & 148 	 & 147  \\
-0.17\% & $ 1''' \leftrightarrow         0      $ &  179 	 & 167 	 & 179  \\
-0.17\% & $ 1''' \leftrightarrow         2      $ &  189 	 & 189 	 & 150  \\
-0.17\% & $ 0    \leftrightarrow         1''    $ &  169 	 & 169 	 & 156  \\
-0.17\% & $ 0    \leftrightarrow         0''    $ &  165 	 & 164 	 & 165  \\
-0.17\% & $ 0    \leftrightarrow         -1'''  $ &  183 	 & 183 	 & 136  \\
 0\%    & $ 1''' \leftrightarrow         1      $ &  154 	 & 154 	 & 153  \\
 0\%    & $ 1''' \leftrightarrow         0      $ &  184 	 & 172 	 & 184  \\
 0\%    & $ 1''' \leftrightarrow         2      $ &  194 	 & 194 	 & 152  \\
 0\%    & $ 0    \leftrightarrow         1''    $ &  174 	 & 174 	 & 161  \\
 0\%    & $ 0    \leftrightarrow         0''    $ &  171 	 & 170 	 & 171  \\
 0\%    & $ 0    \leftrightarrow         -1'''  $ &  189 	 & 189 	 & 138  \\
 0.34\% & $ 1''' \leftrightarrow         1      $ &  164 	 & 164 	 & 161  \\
 0.34\% & $ 1''' \leftrightarrow         0      $ &  185 	 & 174 	 & 185  \\
 0.34\% & $ 1''' \leftrightarrow         2      $ &  200 	 & 200 	 & 154  \\
 0.34\% & $ 0    \leftrightarrow         1''    $ &  179 	 & 179 	 & 168  \\
 0.34\% & $ 0    \leftrightarrow         0''    $ &  180 	 & 179 	 & 180  \\
 0.34\% & $ 0    \leftrightarrow         -1'''  $ &  199 	 & 199 	 & 141  
	\end{tabular}
\end{table}
%%%%%%%%%%%%%%%%%%%%%%%%%%%%%%%%%%%%%%%%%%%%%%%%%%%%%%%%%%%%%%%%%%%%%%%%%%%%%%%%%%%%%%%%%%%5
\begin{figure*}[htbp]
 \includegraphics[width=0.6\textwidth]{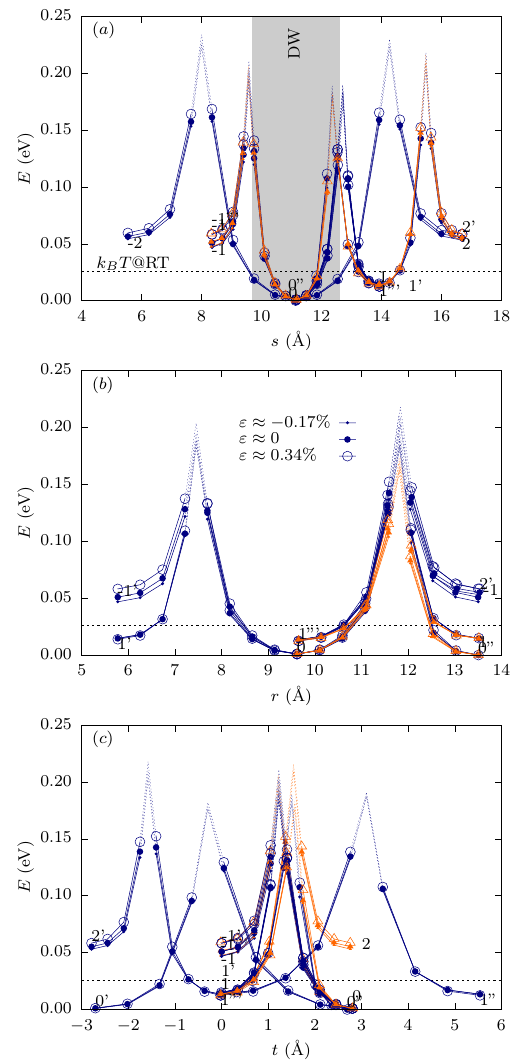}
 \includegraphics[width=0.325\textwidth]{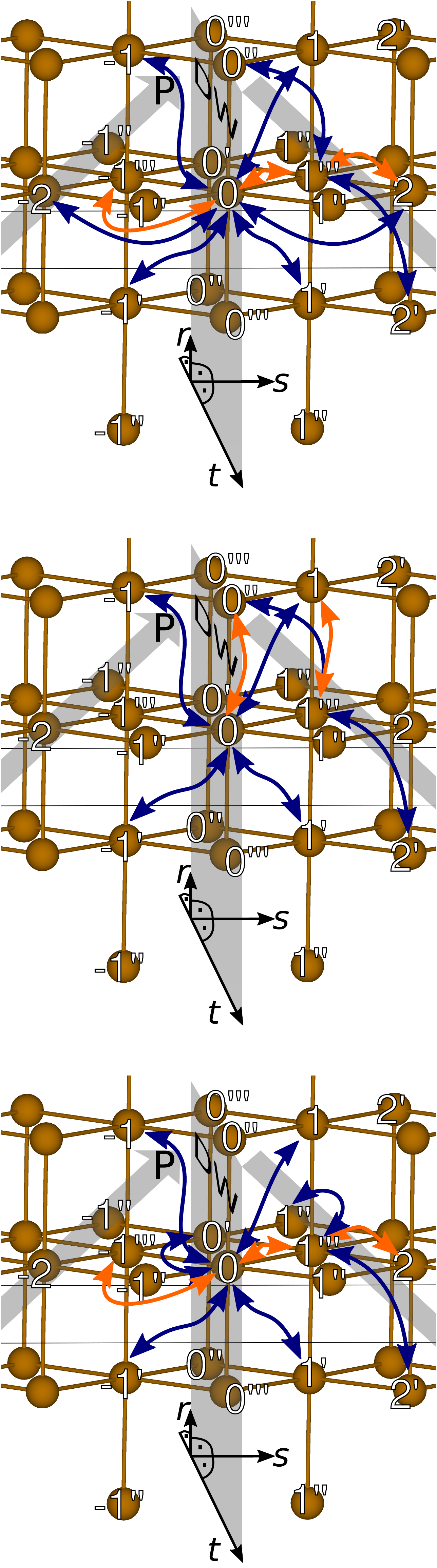}
%	\caption{\label{fig:DW_hops}Top: second-nearest neighbor hops from Fe18, the metastable site. Center: hops from Fe5, the stable site. 
%	Bottom: both.}
	\caption{\label{fig:DW_hops}Energy barriers of polaron hops at the 71\textdegree~domain wall. 
	(a) All hops with a component across the wall, (b) and (c): all hops with components along the wall. 
	Orange triangles: nearest-neighbor, dark-blue circles: second-nearest-neighbor hops. 
	Numbers indicate the Fe positions on the right and in Figure~2(b) in the main text.
	Hop directions were projected on (a) the $s$ coordinate, (b) the  $r$ coordinate, and (c) the $t$ coordinate.}
\end{figure*}
\clearpage
%%%%%%%%%%%%%%%%%%%%%%%%%%%%%%%%%%%%%%
\section{Convergence of energy barriers with respect to the supercell size}
%%%%%%%%%%%%%%%%%%%%%%%%%%%%%%%%%%%%%%
Barriers in bulk were calculated in supercells with 40 and with 80 atoms, respectively. 
\begin{figure*}
        \includegraphics[width=0.49\textwidth]{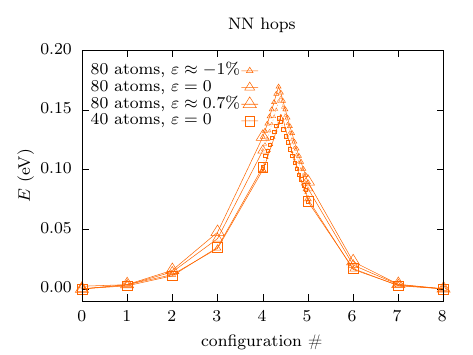}
        \includegraphics[width=0.49\textwidth]{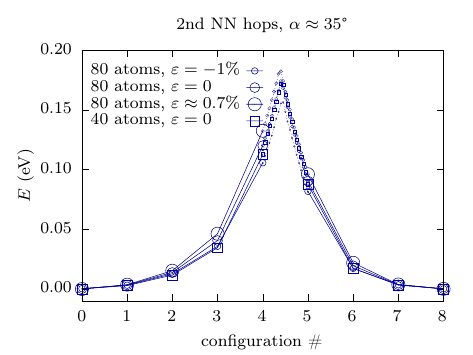}
	\caption{\label{fig:barrier_convergence} \new{Energy barriers for polaron hopping between NN sites and between 2NN sites for supercells with 40 and 80 atoms  
	with different lattice strain $\varepsilon$. For $\varepsilon=0$, the barriers obtained with 40 and 80 atoms differ by about 10\%.}
        }
\end{figure*}
\textbf{Figure}~\ref{fig:barrier_convergence} shows hopping barriers in bulk for different supercell sizes (40 and 80 atoms). 
%For both  supercell sizes, the barriers are close to the thermal energy at room temperature.
The similar results obtained with different supercell sizes indicate that convergence with respect to the supercell size was approximately achieved.

The strain reference (zero strain, $\varepsilon=0$) corresponds to the optimized geometry of bulk BiFeO$_3$ without excess electron.
	Tensile strain, $\varepsilon>0$, corresponds to the optimized geometry of bulk BiFeO$_3$ with an excess electron.

%%%%%%%%%%%%%%%%%%%%%%%%%%%%%%%%%%%%%%
\section{\new{Convergence of energy barriers with respect to the $k$-point mesh}}
%%%%%%%%%%%%%%%%%%%%%%%%%%%%%%%%%%%%%%
Barriers in bulk BiFeO$_3$ were calculated in supercells with 80 atoms, using $k$-point meshes of $2\times 2\times 2$ and $4\times 4\times 4$ $k$-points, respectively. 
\begin{figure*}
        \includegraphics[width=0.49\textwidth]{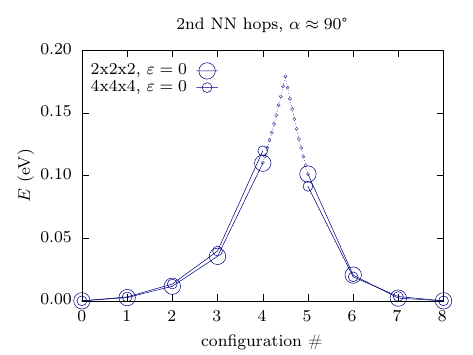}
	\caption{\label{fig:barrier_convergence_k} \new{Energy barriers for polaron hopping between 2NN sites for supercells with 80 atoms  
	and $k$-point meshes of $2\times 2\times 2$ and $4\times 4\times 4$ $k$-points,  respectively, with lattice strain $\varepsilon=0$. 
	 The barriers obtained with the two $k$-point meshes differ by about 10\%.
	}
        }
\end{figure*}
\textbf{Figure}~\ref{fig:barrier_convergence_k} shows hopping barriers in bulk BiFeO$_3$ for two different $k$-point meshes ($1\times 2 \times 2$ and $1\times 4\times 4$ $k$-points). 
The similar results obtained with the two $k$-point meshes indicate that convergence with respect to the $k$-point density was achieved.

%%%%%%%%%%%%%%%%%%%%%%%%%%%%%%%%%%%%%%
\section{\new{Convergence of the energy profile at domain walls with respect to the $k$-point mesh}}
%%%%%%%%%%%%%%%%%%%%%%%%%%%%%%%%%%%%%%
\begin{figure}[htb]
	\includegraphics[width=0.33\textwidth]{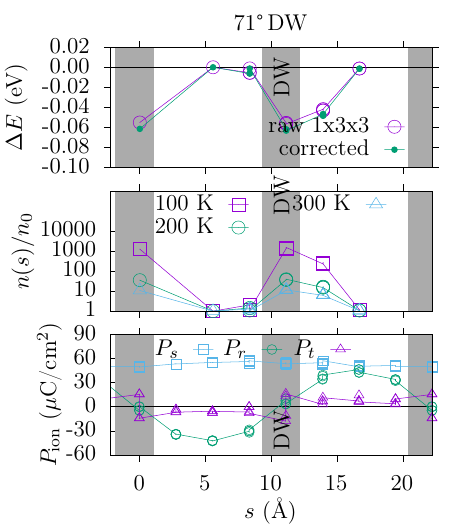}
	\includegraphics[width=0.33\textwidth]{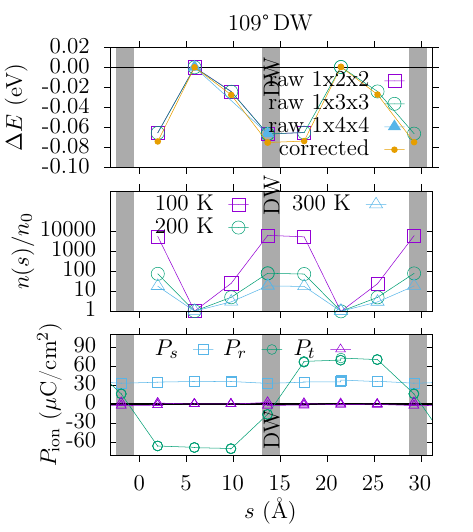}
	\includegraphics[width=0.33\textwidth]{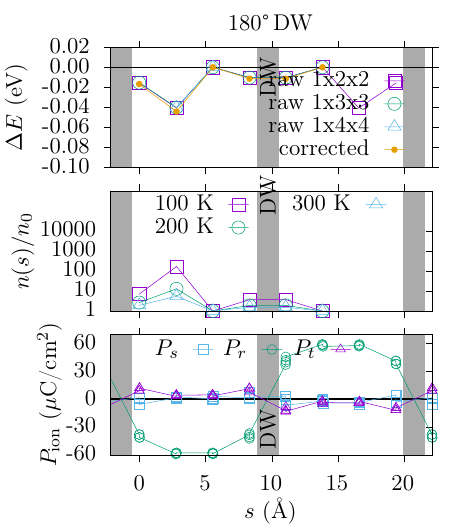}
	\caption{\label{fig:energy_dens_pol_160at}\new{Energy (top), density (center), and polarization (bottom) profile for small electron polarons at the 71\textdegree\ (left), 109\textdegree\ (center), and the 180\textdegree\ DW (right)} obtained in a 160-atom supercell using $k$-point meshes from $1\times 2 \times 2$ to $1\times 4 \times 4$ $k$-points.}
\end{figure}
\new{\textbf{Figure}~\ref{fig:energy_dens_pol_160at} shows the trapping energy, computed polaron density, and ionic polarization profile for small electron polarons at the 71\textdegree, 109\textdegree, and 180\textdegree\ walls 
calculated in 160-atom supercells, using $k$-point meshes from $1\times 2 \times 2$ to $1\times 4 \times 4$ $k$-points. 
The energy profiles obtained with the different $k$-point meshes are nearly identical.}
%%%%%%%%%%%%%%%%%%%%%%%%%%%%%%%%%%%%%%
%    \new{
%\section{Convergence of energy barriers during the nudged-elastic band optimization}
%%%%%%%%%%%%%%%%%%%%%%%%%%%%%%%%%%%%%%%
%\textbf{Figure}~\ref{fig:NEB_energy_vs_iter} shows the calculated energy barrier for the 2NN hop in bulk BiFeO$_3$ with $\alpha\approx$35\textdegree\ during the nudged-elastic band (NEB) optimization. 
%$n_{\mathrm{iter.}}=1$ is the first NEB iteration step, the configurations for $n_{\mathrm{iter.}}=1$ are a linear interpolation between initial (0) and final configuration (8). 
%Here the barrier along the direct path (linear interpolation, $n_{\mathrm{iter.}}=1$) is about nine times that of the minimum-energy path ($n_{\mathrm{iter.}}\gtrsim 10$). 
%\begin{figure}
%    \includegraphics[width=0.49\textwidth]{NEB_energies_vs_iter.pdf}
%    \caption{\label{fig:NEB_energy_vs_iter} \new{Energy barrier for polaron hopping for different iteration steps $n_{\mathrm{iter.}}$ 
%    during the nudged-elastic band (NEB) optimization.}
%        }
%\end{figure}
%}
%%%%%%%%%%%%%%%%%%%%%%%%%%%%%%%%%%%%%%
    \new{
\section{Polaron formation time from molecular dynamics}
%%%%%%%%%%%%%%%%%%%%%%%%%%%%%%%%%%%%%%
\textbf{Figure}~\ref{fig:MD_bulk_10K} shows the total energy, magnetic momenst of Fe atoms, and the temperature during the first nine~femtoseconds of an ab initio molecular dynamics simulation 
for bulk BiFeO$_3$ with an excess electron at \new{an initial temperature of} 10\,K.
The simulation cell contains 80 atoms. The $NVE$ ensemble was chosen. No thermostat was applied.  The time step was set to one~femtosecond.
After about 3 fs, polaron formation sets in, indicated by the energy drop, and the abrupt change in one of the Fe atoms' magnetic moments from the high-spin configuration, 5 times spin down, to 4 times spin down 
(here the computed magnetic moment of about 4\,$\mu_B$ corresponds to the high-spin configuration with 5 parallel spins; due to the choice of cutoff radius for orbital projection the numerical value 
differs from the theoretical value of 5\,$\mu_B$).
\begin{figure}
    \includegraphics[width=0.49\textwidth]{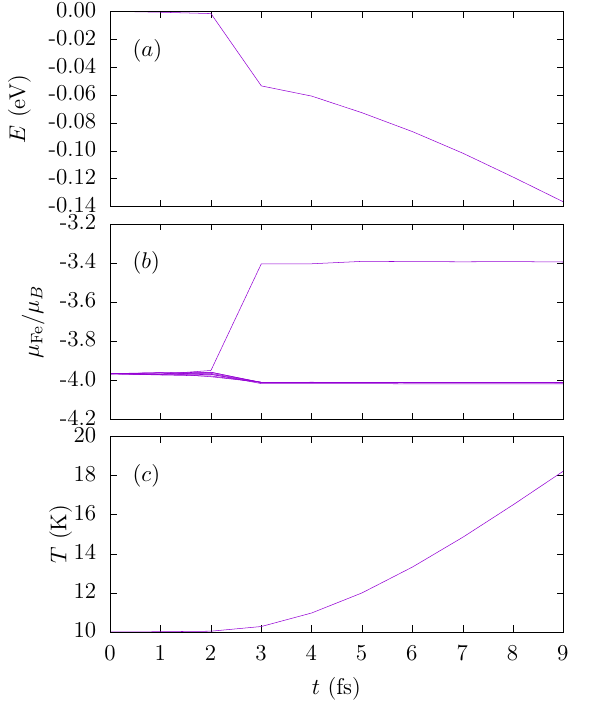}
    \caption{\label{fig:MD_bulk_10K} \new{(a) Total energy, (b) magnetic moments on Fe atoms, and (c) temperature during the first 9 seconds of an ab initio molecular dynamics simulation of bulk BiFeO$_3$ with an excess electron.}
        }
\end{figure}
}
%%%%%%%%%%%%%%%%%%%%%%%%%%%%%%%%%%%%%
%\clearpage
%
%% References
%\medskip
%
%% Use the following code if you wish to generate your bibliography with BibTeX;
%% replace the string "MSP-template" below with the name(s) of
%% the BibTeX data base(s) you want to use.
%% The resulting bibliography-output (the content of the .bbl file)
%% must be pasted back into this file before submission.
%% Please also include your BibTeX data base file(s) in your submission
%% so that we can re-run BibTeX if necessary.
%%
%\bibliographystyle{MSP}
%\bibliography{jabbr,all}
%%\bibliography{/media/koerbel/Elements/Literature/LatexBib/jabbr,/media/koerbel/Elements/Literature/LatexBib/all}

%